\documentclass[12pt]{article}
\usepackage{amsfonts,amssymb,amstext,epsfig}
\usepackage[latin1]{inputenc}
\usepackage[english]{babel}
\usepackage{color}
\usepackage[T1]{fontenc}
\usepackage{makeidx}
\usepackage{graphicx}
\usepackage{subcaption}
\usepackage{mathtools}
\usepackage{braket}
\usepackage{multirow}
\usepackage{amsthm}
\usepackage{systeme}
\usepackage{ragged2e} 
\usepackage{calrsfs}
\usepackage{hyperref}
\newtheorem{thm}{Theorem}[section]
\newtheorem{cor}[thm]{Corollaire}
\newtheorem{lem}[thm]{Lemme}
\newtheorem{pro}[thm]{Proposition}
\newtheorem{dfn}[thm]{Definition}
\newtheorem{rmk}[thm]{Remark}
\newtheorem{expl}[thm]{Exemple}

\newcommand{\ba}{\begin{array}}
	\newcommand{\ea}{\end{array}}
\newcommand{\MeijerG}[5]{G^{#1}_{#2}
	\def\deq{\stackrel{\mathrm{def}}{=}}
	\left(#5\middle\vert\
	\begin{smallmatrix} #3 \\ \\ \\ #4	
	\end{smallmatrix} 
	\right) }

\def\dessous#1\sous#2{\mathrel{\mathop{\kern0pt#2}\limits_{#1}}}

\newcommand{\beq}{\begin{eqnarray}}
\newcommand{\eeq}{\end{eqnarray}}
\newcommand{\bpro}{\begin{pro}}
\newcommand{\epro}{\end{pro}}
\newcommand{\blem}{\begin{lem}}
\newcommand{\elem}{\end{lem}}
\newcommand{\bdfn}{\begin{dfn}}
\newcommand{\edfn}{\end{dfn}}
\newcommand{\bcor}{\begin{cor}}
\newcommand{\ecor}{\end{cor}}
\newcommand{\bthm}{\begin{thm}}
\newcommand{\ethm}{\end{thm}}
\newcommand{\bex}{\begin{expl}}
\newcommand{\eex}{\end{expl}}
\newcommand{\brmk}{\begin{rmk}}
\newcommand{\ermk}{\end{rmk}}
\newcommand{\benum}{\begin{enumerate}}
\newcommand{\eenum}{\end{enumerate}}
\newcommand{\bitem}{\begin{itemize}}
\newcommand{\eitem}{\end{itemize}}
\def\deq{\stackrel{\mathrm{def}}{=}}
\begin{document}
\begin{center}
{\Large\bf {Constructing Barut-Girardello coherent states for the isotonic oscillator in the DOOT approach }}

 \vspace{0.5cm}

 Messan M\'{e}dard  Akouetegan$^{1}$, Isiaka Aremua$^{1,2}$
 and Mahouton Norbert Hounkonnou$^{2}$
\vspace{0.5cm}

 $^{1}${\em
Laboratoire de Physique des Mat\'{e}riaux et des Composants \`{a} Semi-Conducteurs, Facult\'{e} Des Sciences (FDS),
D\'{e}partement de Physique, Universit\'{e} de Lom\'{e} (UL), 01 B.P. 1515 Lom\'{e} 01, Togo}\\
$^{2}${\em University of Abomey-Calavi, International Chair in Mathematical Physics and Applications (ICMPA), 072 B.P. 050 Cotonou, Benin} \\
E-mail:  akoueteganmessanmedard@gmail.com, claudisak@gmail.com, norbert.hounkonnou@cipma.uac.bj
%  $^{2}${\em
% The Abdus Salam International Centre for Theoretical Physics (ICTP),
% Strada Costiera 11, I-34151 Trieste Italy. E-mail: laure.gouba@gmail.com}

\vspace{1.0cm}

\today

%\clearpage
\begin{abstract}
\noindent
In this work, we study the quantum system of the isotonic oscillator from the perspective of the diagonal operator ordering technique (DOOT). Within this framework, we construct the associated Barut-Girardello and Gazeau-Klauder coherent states. We examine  their mathematical properties using reproducing kernels and compute the expectation values of observables that characterize the system and its relevant physical features. Further, we perform the quantization of  main classical variables in the complex plane. Then, by exploring the thermal behavior of the physical system in the constructed coherent states, we analyze the properties of mixed states described by a canonical density operator. We also obtain the corresponding Glauber-Sudarshan $P$-representation.
\end{abstract}

\end{center}


\setcounter{footnote}{0}

\section{Introduction}
Various kinds of quantum optical systems \cite{barut-girardello, perelomov} have recently been explored by many researchers to investigate the nonclassical properties of light in quantum optics \cite{zhang-feng, gilmore}. The $su(1,1)$ Lie algebra and Barut-Girardello coherent states (BGCSs), for their mathematical and physical insights,  have been extensively studied in connection with the thermal and statistical behavior of quantum models. Examples include physical systems governed by a Hamiltonian operator in two-dimensional space, such as spinless charged particles subject to a perpendicular magnetic field and coupled with a harmonic potential, the pseudoharmonic oscillator, the isotonic oscillator, and others. This is evidenced by numerous recent publications \cite{aremuaromp,Wangetal, Dodonov, Thirulogasanthar-Saad, Ranada, Popov2001, brif-aryeh} and the references therein. There is a wealth of significant books and review papers on coherent states (CSs), their generalizations, and applications \cite{perelomov,zhang-feng,Klauder-Skagerstam,Dodonov,Gazeau,ali-antoine-gazeau}, since their introduction by Schr\"odinger in 1926 \cite{Schrodinger}.

A family of vectors in a Hilbert space, which may be finite or countably infinite, describing the quantum system can be considered as a family of CSs if it satisfies a minimum set of criteria: normalizability, continuity, and a resolution of the identity with a positive weight function \cite{Klauder-Skagerstam}. CSs can be defined over complex domains in the Hilbert space $\mathfrak{H} = span \{\phi_{m}, m \in \mathbb N \}$, as ket vectors labeled by a complex number \cite{ali-antoine-gazeau}:

{\small{
		\begin{equation}{\label{covcs}}
		|z\rangle  = (\mathcal{N}(|z|^2))^{-\frac{1}{2}}\sum_{m=0}^{\infty}\frac{z^{m}}{\sqrt{\rho(m)}}|\phi_{m}\rangle,
		\qquad \mathcal{N}(|z|^2)  = \sum_{m=0}^{\infty}\frac{(|z|^2)^m}{\rho(m)}
		\end{equation}
}}
% 		{\color{blue}
$z = re^{i\theta}, 0 \leq r \leq  \infty, 0\leq \theta \leq 2\pi, $
% 		}
where $\{\rho(m)\}_{m=0}^{\infty}$ is a sequence of nonzero positive numbers that determines the internal structure of the CSs. This sequence is chosen to ensure the convergence of the sum in a nonempty open subset $\mathcal D$ of the complex plane, and $\mathcal N(|z|^2)$ is the normalization factor ensuring that $\langle z|z\rangle  = 1$.

The resolution of the identity is given by
{\small{
		\begin{equation*}{\label{rescovcs}}
		\int_{\mathcal{D}}|z\rangle \langle z| d\nu = I_{\mathfrak{H}},
		\end{equation*} 
}}
where $d\nu$ is an appropriate chosen measure and $I_{\mathfrak{H}}$ the identity operator on the Hilbert space $\mathfrak H$.

In recent years, the diagonal operator ordering technique (DOOT), denoted in calculations by the symbol $\# \#$, was introduced by D. Popov {\it et al.} \cite{Popov-popov-miodrag, Popov-Negrea-Popov}  to generalize the operatorial method based on integration within a normally ordered product of operators (IWOP). This technique, originally formulated for the creation and annihilation operators of the harmonic oscillator coherent states (CSs), enables the construction of generalized hypergeometric coherent states (GHCSs) and has been extended to CSs of other types of oscillators, such as the Morse oscillator \cite{Popov-Dong-miodrag} and its related Barut-Girardello coherent states (BGCSs), including discussions of their thermal states.

The IWOP technique itself was proposed and developed by Hong-yi Fan and coauthors for Bose operators, particularly in relation to CSs of the one-dimensional harmonic oscillator (HO-1D) (\cite{Fan H-Y,Fan H-Y_al2,Fan H-Y_al,Fan H-Y_al4}, as well as further developments referenced therein). Within the context of the Meixner oscillator \cite{Popov-new}, DOOT has been applied to pairs of raising and lowering operators $A^+$ and $A^-$ for analyzing the Mandel parameter and statistical properties in pure and mixed (thermal) states of Barut-Girardello and Klauder-Perelomov CSs. DOOT has also proven valuable for investigating the properties of $k$-CSs \cite{Popov-k} and $k$-generalized hypergeometric BGCSs \cite{Popov-thermal}. Furthermore, in \cite{Popov-negrea}, the DOOT approach has been  successfully used to study the properties of Barut-Girardello, Klauder-Perelomov, and Gazeau-Klauder CSs, as well as to deduce integrals involving hypergeometric and Meijer functions. More recently, in \cite{Popov-consideration}, the DOOT technique, as a generalization of IWOP, has been employed to study two types of nonlinear CSs (NCSs) via properties of generalized hypergeometric and $G$-Meijer functions.

Motivated by these recent developments in the DOOT procedure, and noting that this technique is not only applicable to CSs of the one-dimensional harmonic oscillator (HO-1D) but is also effective for the analysis of the pseudoharmonic oscillator and its associated BGCSs  \cite{Popov-popov-miodrag}, as well as generalizable to other types of NCSs  \cite{Popov-consideration}, we introduce and apply in this work the rules for DOOT calculus.

The structure of the paper is as follows. Section 2 presents the quantum system of interest, emphasizing its underlying $\mathfrak{su}(1,1)$ Lie algebraic structure. In Section 3, we introduce the Fock Hilbert space basis necessary for the application of various rules of the DOOT technique. This includes an explicit representation of the $\mathfrak{su}(1,1)$ generators and the construction of the normalized vacuum projector within the DOOT formalism. Subsequently, we construct the associated Barut-Girardello coherent states (BGCSs) for the isotonic oscillator, categorizing them into even and odd classes within the DOOT framework on the established Fock Hilbert space. Additionally, the corresponding Gazeau-Klauder coherent states (GKCSs) are derived in the DOOT approach, beginning with the appropriate Fock Hilbert space constructed from the eigenvectors of the quantum Hamiltonian.

Section 4 delves into the mathematical properties of each family of constructed coherent states, examining their completeness via the inverse Mellin transform and deriving suitable weight functions. The reproducing kernel property is then discussed for each class of coherent states. Section 5 is devoted to the evaluation of the expectation values of the $\mathfrak{su}(1,1)$ Lie algebra generators in the various coherent state bases. In Section 6, we analyze the statistical behavior of the quantum system by formulating the relevant density operator in each coherent state representation. The $Q$-Husimi function is subsequently obtained, followed by the $P$-diagonal representation of each corresponding density operator. Finally, Section 7 summarizes the main results and offers concluding remarks.

\section{The isotonic oscillator quantum Hamiltonian and the underlying $\mathfrak{su}(1,1)$ Lie algebra}
The quantum system of the model known as the isotonic oscillator is obtained through a deformation of the one-dimensional harmonic oscillator described by the classical Hamiltonian
\begin{equation}\label{Isotonic1}
H(x,p_{x})=\frac{p^{2}_{x}}{2m}+\frac{1}{2}m\omega^{2}x^{2}
\end{equation}
where $(x,p_x)\in \mathbb{R}^{+}\times\mathbb{R}$ implying $x>0$,and where the canonical Poisson bracket is given by $\{x,p_x\}=1$. By introducing the variable $d_x = p_x x$, called the dilation variable, together with the variables $x$, and $d_x$ referred to as affine variables, which satisfy the Poisson bracket relation $\{x,d_x\}=x$ \cite{Klauder2012,Gouba2021}, the Hamiltonian in (\ref{Isotonic1}) takes the form
\begin{equation}\label{Isotonic11}
H(x,d_x)=\frac{d_{x}(x^{-2})d_{x}}{2mx^{2}}+\frac{1}{2}\omega^{2}x^{2}.
\end{equation}
Within the framework of affine quantization, the classical affine variables are promoted to operators acting on the wave function $\psi(x,t)$ as
\begin{eqnarray}
&&	x \mapsto \hat{x} :\quad\left\{\hat{x}\psi(x,t)=x\psi(x,t),\quad x>0\right\} \nonumber\\
&& d_x \mapsto  \hat{d_x}:\quad\left\{\hat{d_x}\psi(x,t)=-i\hbar\left(x\frac{\partial}{\partial x}+\frac{1}{2}\right)\psi(x,t)\right\},\nonumber
\end{eqnarray}
which satisfy the following commutation relation:
\begin{equation*}
[\hat{x},\hat{d_x}]=i\hbar\hat{x}.
\end{equation*}
Consequently, the Hamiltonian operator of the harmonic oscillator takes the form
\begin{equation*}
\hat{H}(\hat{x},\hat{d_x})=\frac{\hat{d_x} (\hat{x})^{-2}\hat{d_x}}{2m}+\frac{1}{2}m\omega^{2}\hat{x}^{2}
\end{equation*}
leading to the Schr\"odinger equation
\begin{equation*}
i\hbar\frac{\partial}{\partial t}\psi(x,t)=\hat{H}\psi(x,t).
\end{equation*}
The wave functions $\psi(x,t)$, defined in \cite{Klauder2012,Gouba2021}, are given by
\begin{equation*}
\psi(x,t)=\phi(x)e^{-iEt/\hbar},
\end{equation*}
and are normalized according to the condition
\begin{equation*}
\int_{0}^{\infty}|\psi(x,t)|^{2},dx=1.
\end{equation*}
Substituting this expression into the Schr\"odinger equation leads to the time-independent eigenvalue equation
\begin{equation}\label{Isotonic2}
\hat{H}(\hat{x},\hat{d_x})\phi(x)=E\phi(x).
\end{equation}
By applying the action of the operators $\hat{x}$ and $\hat{d_x}$, one successively obtains
\begin{equation*}
\hat{d}_x\phi(x)
=-i\hbar\left(x\phi'(x)+\frac{1}{2}\phi(x)\right),
\end{equation*}

\begin{equation*}
\hat{x}^{-2}\hat{d}_x\phi(x)
=-i\hbar\left(\frac{1}{x}\phi'(x)+\frac{1}{2x^{2}}\phi(x)\right),
\end{equation*}
\begin{equation*}
\hat{d}_x\hat{x}^{-2}\hat{d}_x\phi(x)
=-\hbar^{2}\left(
\phi''(x)-\frac{3}{4x^{2}}\phi(x)
\right).
\end{equation*}
The eigenvalue equation (\ref{Isotonic2}) can therefore be written in the following explicit form:
\begin{equation}
\left[-\frac{\hbar^{2}}{2m}\frac{d^{2}}{dx^{2}}+\frac{3\hbar^{2}}{8mx^{2}}+\frac{1}{2}m\omega^{2}x^{2}\right]\phi(x)=E\phi(x).
\end{equation}
By choosing $m=1$, $\omega=1$, and $\hbar=1$ to simplify the equation, one finally obtains
\begin{equation}
\left[-\frac{1}{2}\frac{d^{2}}{dx^{2}}+\frac{1}{2}x^{2}+\frac{1}{2}\frac{3}{4x^{2}}\right]\phi(x)=E\phi(x).
\end{equation}
The resulting  Schr\"odinger equation corresponds to a Hamiltonian that is a particular case of the isotonic oscillator Hamiltonian, with the parameter $A=\frac{3}{4}$.

In this work, we consider the one-dimensional isotonic oscillator, whose generalized Hamiltonian is given by \cite{Thirulogasanthar-Saad, Ranada,Hall-Saad}
\begin{equation}\label{eq2}
H=-\frac{1}{2}\frac{d^2}{dx^2} + \frac{1}{2}x^2 + \frac{1}{2}\frac{A}{x^2},\qquad(A=g(g+1)\ge 0),
\end{equation}
acting in the Hilbert space $L^2(\mathbb{R}^+, dx)$ and the eigenfunctions $\Psi$ $\in$ $L^2 (0,\infty)$ satisfy the Dirichlet boundary condition; $\Psi(0) = 0$. The Hamiltonian is the generalization of the harmonic oscillator Hamiltonian in 3-dimension where the generalization lies in the parameter A ranging over $[0,\infty)$ instead of the angular momentum quantum numbers $l = 0, 1, 2, \dots$. The Hamiltonian $H$ admits exact solutions \cite{mandel-wolf}
\begin{equation*}
\Phi^{\gamma}_m(x)=\sqrt{\frac{2(\gamma)_m}{
		m!\Gamma(\gamma)}}x^{\gamma-\frac{1}{2}}e^{-\frac{1}{2}x^2}\, _1F_1(-m;\gamma;x^2),\qquad m=0,1,2\dots
\end{equation*}
where $\gamma=1+\frac{1}{2}\sqrt{1+4g(g+1)}$ and the exact eigenvalues are
given by
\begin{equation*}
E_m=2(2m+\gamma), \qquad  m = 0,1, 2,\dots.
\end{equation*}
It was also shown that
\begin{equation*}
\int_{0}^{\infty}\Phi^{\gamma}_m(x)\Phi^{\gamma}_n(x)dx=\delta_{m,n}
\end{equation*}
and the collection of vectors $\left\{|\Phi_m^{\gamma}\rangle\right\}_{m = 0}^{\infty}$ form an orthonormal basis for the Hilbert space $\mathfrak{H}=L^2(\mathbb{R}^+, dx)$.

Consider the quantum Hamiltonian $H$ (\ref{eq2}) describing the isotonic oscillator, and introduce the following  generators 
\begin{equation*}
W_1=\frac{1}{2}x^2,\qquad W_2=-\frac{i}{2}\left(x\frac{d}{dx}-\frac{d}{dx}x\right),\qquad W_3=-\frac{1}{2}\frac{d^2}{dx^2} + \frac{1}{2}\frac{g(g+1)}{x^2}
\end{equation*}
obeying  the following commutation relations
\begin{equation}
[W_1,W_2]=iW_1,\qquad [W_2,W_3]=iW_3,\qquad [W_1,W_3]=2iW_2.
\end{equation}
In addition, consider the set of operators $K_i (i=1,2,3)$ given by
\begin{eqnarray}
K_1&=&\frac{1}{2}(W_3-W_1)=\frac{1}{2}\left[-\frac{1}{2}\frac{d^2}{dx^2} - \frac{1}{2}x^2 + \frac{1}{2}\frac{g(g+1)}{x^2}\right],\nonumber\\
K_2&=&W_2=-\frac{i}{2}\left(x\frac{d}{dx}-\frac{d}{dx}x\right),\nonumber\\
K_3&=&\frac{1}{2}(W_3+W_1)=\frac{1}{2}\left[-\frac{1}{2}\frac{d^2}{dx^2} + \frac{1}{2}x^2 + \frac{1}{2}\frac{g(g+1)}{x^2}\right],\nonumber
\end{eqnarray}
satisfying
\begin{equation}
[K_1,K_2]=-iK_3,\qquad[K_2,K_3]=iK_1,\qquad[K_3,K_1]=iK_2.
\end{equation}
The above operators are related to  the Lie algebra $\mathfrak{su}(1, 1)$ \cite{gilmore} generators   corresponding to the $SU(1, 1)$ group,
spanned by  $\{K_+, K_-, K_0\}$  and verifying 
\begin{equation}
K_+=\frac{1}{2}(K_1-iK_2),\qquad K_-=\frac{1}{2}(K_1+iK_2), \qquad\quad 2K_0=K_3.
\end{equation}
%Therefore
There results the following representation of the Lie algebra 
$\mathfrak{su}(1,1)$ in the Hilbert space $\mathfrak H = span\{\ket{n,\gamma}\}_{n= 0}^{\infty}$  given by
\begin{eqnarray}\label{gene000}
K_0\ket{n,\gamma}&=&(2n+\gamma)\ket{n,\gamma},\cr
K_-\ket{n,\gamma}&=&\sqrt{2n(n+\gamma-1)}\ket{n-1,\gamma},\cr
K_+\ket{n,\gamma}&=&\sqrt{2(n+1)(n+\gamma)}\ket{n+1,\gamma}, 
\end{eqnarray} 
where $\gamma=1+\frac{1}{2}\sqrt{1+4g(g+1)}$ is Bargmann index.

Consquently, we get
\begin{equation*}
\langle n,\gamma|K_+K_-\ket{n,\gamma}=2n(n+\gamma-1),
\end{equation*}
and the commutation of annihilation and creation operators
\begin{eqnarray}
[K_-,K_+]&=&2K_0.
\end{eqnarray}
\section{The DOOT approach  and related  coherent states}
Let us begin this section by recalling the main principles of the diagonal operator ordering technique (DOOT) \cite{Popov-CSS}. In the remainder of this work, we apply the DOOT framework to the isotonic oscillator system, focusing on the $\mathfrak{su}(1,1)$ Lie algebra generators and the associated Barut-Girardello coherent states (BGCSs), as well as the Gazeau-Klauder coherent states (GKCSs), defined on the corresponding Fock Hilbert spaces.

%Let us start this section by recalling the main rules of DOOT technique \cite{Popov-CSS}. In the rest of this work we apply the DOOT to the isotonic oscillator  system, regarding the $\mathfrak{su}(1,1)$ Lie  algebra generators and related BGCSs, and also GKCSs, on the corresponding Fock Hilbert spaces.
\subsection{The $\mathfrak{su}(1,1)$ generators representation in the DOOT framework}
From the representation of the $\mathfrak{su}(1,1)$ generators on the  Hilbert space $\mathfrak H = span\{\ket{n,\gamma}\}_{n= 0}^{\infty}$ given by Eqs.(\ref{gene000}), the Fock basis vectors $\ket{n,\gamma}$ and their duals $\bra{n,\gamma}$  can be constructed by successive applications of the raising operator $K_+$ on the vacuum state $\ket{0,\gamma}$. One finds
\begin{equation*}
(K_+)^n|0,\gamma\rangle = \sqrt{4^n\left(\frac{\gamma}{2}+1\right)_n}|n,\gamma\rangle,
\end{equation*}
where  $\left(\frac{\gamma}{2}+1\right)_n$ denotes the Pochhammer symbol, defined by \begin{equation*}
\left(\frac{\gamma}{2}+1\right)_n  = \left(\frac{\gamma}{2}+1\right) \left(\frac{\gamma}{2}+2\right)\left(\frac{\gamma}{2}+3\right)\cdots \left(\frac{\gamma}{2}+n\right). 
\end{equation*}
Consequently, the ket and bra vectors take the explicit operator form
\begin{equation}\label{DOOTketbra000}
|n,\gamma\rangle = \frac{(K_+)^n}{\sqrt{4^n\left(\frac{\gamma}{2}+1\right)_n}}|0,\gamma\rangle, \qquad 
\langle n,\gamma|=\langle 0,\gamma|\frac{(K_-)^n}{\sqrt{4^n\left(\frac{\gamma}{2}+1\right)_n}}.
\end{equation}
Combinning the two relations in Eq.(\ref{DOOTketbra000}), and using the completeness of the Fock basis in $\mathfrak H = span\{\ket{n,\gamma}\}_{n= 0}^{\infty}$,  the resolution of the identity operator $\mathbb{I_{\mathfrak H}}$ can be written in the diagonal operator-ordered form
\begin{equation}\label{Over1}
\sum_{n=0}^{\infty}|n,\gamma\rangle\langle n,\gamma| = 
\mathbb{I_{\mathfrak H}}=\sum_{n=0}^{\infty}\frac{(K_+K_-)^n}{4^n\left(\frac{\gamma}{2}+1\right)_n}|0,\gamma\rangle\langle 0,\gamma|.
\end{equation}
The Barut-Girardello coherent states (BGCSs) associated with the isotonic oscillator are defined as eigenstates of the lowering generator $K_-$ of the $\mathfrak{su}(1,1)$ Lie algebra. Within the DOOT framework, these states naturally split into two distinct classes, namely even and odd coherent states.

Before proceeding with their explicit construction, it is convenient to introduce the following subspaces of $\mathfrak{H}$, as discussed in \cite{Dehghani2015,wunsche}. These subspaces are spanned by the even and odd Fock vectors $|2n,\gamma\rangle$ and $|2n+1,\gamma\rangle$, respectively, and denoted $\mathfrak{H}_e$ and $\mathfrak{H}_o$:
\begin{equation} 
\mathfrak{H}_e :=span\left\{|2n,\gamma\rangle : \langle 2m,\gamma|2n,\gamma\rangle   = \delta_{nm}, \; \sum_{n=0}^{\infty} |2n,\gamma\rangle\langle 2n,\gamma| = \mathbb{I}_{\mathfrak{H}_{e}}
\right\},
\end{equation}

% \begin{eqnarray}
% \mathfrak{H}_o &:=& span\left\{|2n+1,\gamma\rangle : \langle 2m+1,\gamma|2n+1,\gamma\rangle   = \delta_{nm}, \; \sum_{n=0}^{\infty} |2n+1,\gamma\rangle\right.\cr &&\left. \langle 2n+1,\gamma| = \mathbb{I}_{\mathfrak{H}_{o}}
% \right\}
% \end{eqnarray}
{\small
	\begin{equation}
	\mathfrak{H}_o := span\left\{|2n+1,\gamma\rangle : \langle 2m+1,\gamma|2n+1,\gamma\rangle   = \delta_{nm}, \; \sum_{n=0}^{\infty} |2n+1,\gamma\rangle \langle 2n+1,\gamma| = \mathbb{I}_{\mathfrak{H}_{o}}
	\right\},
	\end{equation}
	where 
	% the Hilbert space $\mathfrak H = span\{|n,\gamma\rangle\}_{n= 0}^{\infty}$ is such that
	$\displaystyle  \mathfrak{H} =  \mathfrak{H}_{e} \oplus  \mathfrak{H}_{o}$ with $\displaystyle \mathbb{I}_{\mathfrak{H}}= \displaystyle\mathbb{I}_{\mathfrak{H}_{e}} \oplus \mathbb{I}_{\mathfrak{H}_{o}}$, $\mathbb{I}_{\mathfrak{H}_{e}}$ and $\mathbb{I}_{\mathfrak{H}_{o}}$ being the corresponding identity operators,  considered in the sequel of the work.}

For this purpose, we first derive the projector onto the vacuum state, denoted by $|0,\gamma\rangle\langle 0,\gamma|$ associated with the even and odd Hilbert subspaces, respectively, starting from (\ref{Over1})
\begin{enumerate}
	\item [(a)] Even coherent states.
	
	Restricting (\ref{Over1}) to the even sector and applying the DOOT prescription, one obtains
	\begin{eqnarray}
	\mathbb{I_{\mathfrak H}}&=&\sum_{n=0}^{\infty}\frac{(K_+K_-)^{2n}}{4^{2n}\left(\frac{\gamma}{2}+1\right)_{2n}}|0,\gamma\rangle\langle 0,\gamma|\cr
	% &=&\sum_{p=0}^{\infty}\frac{(K_+K_-)^{p}}{4^{p}\left(\frac{\gamma}{2}+1\right)_{p}}|0,\gamma\rangle\langle 0,\gamma|\cr
	&=& _1F_1\left(1;\frac{\gamma}{2}+1;\frac{K_+K_-}{4}\right)|0,\gamma\rangle\langle 0,\gamma|, 
	\end{eqnarray}
	where $_1F_1$ denotes the confluent hypergeometric function \cite{magnus-oberhettinger-soni}, defined by
	\begin{equation*}
	_1F_1\left(1;\frac{\gamma}{2}+1;z\right) = \sum_{m=0}^{\infty}\frac{z^m}{m! \left(\frac{\gamma}{2}+1\right)_m}.
	\end{equation*}
	Consequently, the vacuum projector in the even sector can be written in the diagonal operator-ordered form
	\begin{equation}\label{vacpro000}
	|0,\gamma\rangle\langle 0,\gamma|=\#\frac{\mathbb{I_{\mathfrak H}}}{_1F_1\left(1;\frac{\gamma}{2}+1;\frac{K_+K_-}{4}\right)}\#.
	\end{equation}
	\item[(b)] Odd coherent states.
	
	Following the same procedure for the odd sector yields 
	\begin{eqnarray}
	\mathbb{I_{\mathfrak H}}&=&\sum_{n=0}^{\infty}\frac{(K_+K_-)^{2n+1}}{4^{2n+1}\left(\frac{\gamma}{2}+1\right)_{2n+1}}|0,\gamma\rangle\langle 0,\gamma|\cr
	% &=&\sum_{i=1}^{\infty}\frac{(K_+K_-)^{p}}{4^{i}\left(\frac{\gamma}{2}+1\right)_{p}}|0,\gamma\rangle\langle 0,\gamma|\cr
	&=&\left( _1F_1\left(1;\frac{\gamma}{2}+1;\frac{K_+K_-}{4}\right)-1\right)|0,\gamma\rangle\langle 0,\gamma|, \nonumber
	\end{eqnarray}
	which leads to the diagonal operator-ordered expression                                                                                             
	\begin{equation}\label{vacpro001}                                                |0,\gamma\rangle\langle 0,\gamma|=\#\frac{\mathbb{I_{\mathfrak H}}}{\left( _1F_1\left(1;\frac{\gamma}{2}+1;\frac{K_+K_-}{4}\right)-1\right)}\#.
	\end{equation}
	As a direct consequence, the expectation values {\small$\langle 0,\gamma|(K_+K_-)^{2n}|0,\gamma\rangle$ and $\langle 0,\gamma|(K_+K_-)^{2n+1}|0,\gamma\rangle$} can be readily evaluated, as shown below : 
	% 
	% The overlap $\langle 2n,\gamma|2n,\gamma\rangle$ show the expression of exceptation value $\langle 0,\gamma|(K_+K_-)^{2n}|0,\gamma\rangle$ as follows:
	\begin{equation}
	% \langle 2n,\gamma|2n,\gamma\rangle&=&\langle 0,\gamma|\frac{(K_+K_-)^{2n}}{4^{2n}\left(\frac{\gamma}{2}+1\right)_{2n}}|0,\gamma\rangle\cr
	% 1&=&\frac{1}{4^{2n}\left(\frac{\gamma}{2}+1\right)_{2n}}\langle 0,\gamma|(K_+K_-)^{2n}|0,\gamma\rangle\cr
	\langle 0,\gamma|(K_+K_-)^{2n}|0,\gamma\rangle = 
	4^{2n}\left(\frac{\gamma}{2}+1\right)_{2n}, 
	\end{equation}
	% Then, the calculations of the exceptation values $\langle 0,\gamma|(K_+K_-)^{2n+1}|0,\gamma\rangle$  is obtained such as one performed for even coherent states follows as:
	\begin{equation}
	% \langle 2n+1,\gamma|2n+1,\gamma\rangle&=&\langle 0,\gamma|\frac{(K_+K_-)^{2n+1}}{4^{2n+1}\left(\frac{\gamma}{2}+1\right)_{2n+1}}|0,\gamma\rangle\cr
	% 1&=&\frac{1}{4^{2n+1}\left(\frac{\gamma}{2}+1\right)_{2n+1}}\langle 0,\gamma|(K_+K_-)^{2n+1}|0,\gamma\rangle\cr
	\langle 0,\gamma|(K_+K_-)^{2n+1}|0,\gamma\rangle = 4^{2n+1}\left(\frac{\gamma}{2}+1\right)_{2n+1}, 
	\end{equation}
\end{enumerate}
respectively. 
\subsection{Barut-Giradello coherent states}
\subsubsection{Construction}

Using the above results, the Barut-Girardello coherent states (BGCSs) associated with the isotonic oscillator are constructed, within the DOOT framework, in the form of two distinct families: even and odd coherent states, denoted by $\ket{z,\gamma}_e$ and $\ket{z,\gamma}_o$,   respectively. Here $z$ is an arbitrary complex parameter written as  $z = \rho e^{i \theta}, 0 \leq \rho < \infty, 0 \leq \theta < 2\pi$. These states are explicitly given as follows :
% 
% {\small\begin{eqnarray}
% \ket{z,\gamma}_e&=&\frac{1}{\sqrt{_1F_1\left(1;\frac{\gamma}{2}+1;\frac{|z|^2}{4}\right)}}\sum_{n=0}^{\infty}\frac{z^{2n}}{\sqrt{4^{2n}\left(\frac{\gamma}{2}+1\right)_{2n}}}\ket{2n,\gamma}, \\
% \ket{z,\gamma}_o&=&\frac{1}{\sqrt{_1F_1\left(1;\frac{\gamma}{2}+1;\frac{|z|^2}{4}\right)}-1}\sum_{n=0}^{\infty}\frac{z^{2n+1}}{\sqrt{4^{2n+1}\left(\frac{\gamma}{2}+1\right)_{2n+1}}}\ket{2n+1,\gamma}. 
% \end{eqnarray}}
% The BGCSs were performed in the DOOT frame as follows:
\begin{enumerate}
	\item [(i)] Even Barut-Giradello CSs. 
	
	The even BGCSs read
	\begin{eqnarray}
	\ket{z,\gamma}_e&=&\frac{1}{\sqrt{_1F_1\left(1;\frac{\gamma}{2}+1;\frac{|z|^2}{4}\right)}}\#_1F_1\left(1;\frac{\gamma}{2}+1;\frac{zK_+}{4}\right)\#\ket{0,\gamma}.\nonumber
	% 	\\
	% 	_e\langle z,\gamma|&=&\langle 0,\gamma|\frac{1}{\sqrt{_1F_1\left(1;\frac{\gamma}{2}+1;\frac{|z|^2}{4}\right)}}\#_1F_1\left(1;\frac{\gamma}{2}+1;\frac{\overline{z}K_-}{4}\right)\#
	\end{eqnarray}
	The corresponding projector onto the state $\ket{z,\gamma}_e$ is obtained as
	\begin{equation}
	\ket{z,\gamma}_e\,_e\langle z,\gamma|
	% &=&\frac{\#_1F_1\left(1;\frac{\gamma}{2}+1;\frac{zK_+}{4}\right)\#}{\sqrt{_1F_1\left(1;\frac{\gamma}{2}+1;\frac{|z|^2}{4}\right)}}\ket{0,\gamma}\langle 0,\gamma|\frac{\#_1F_1\left(1;\frac{\gamma}{2}+1;\frac{\overline{z}K_-}{4}\right)\#}{\sqrt{_1F_1\left(1;\frac{\gamma}{2}+1;\frac{|z|^2}{4}\right)}}\cr
	=\frac{1}{_1F_1\left(1;\frac{\gamma}{2}+1;\frac{|z|^2}{4}\right)}\#\frac{_1F_1\left(1;\frac{\gamma}{2}+1;\frac{\overline{z}K_-}{4}\right)\,_1F_1\left(1;\frac{\gamma}{2}+1;\frac{zK_+}{4}\right)}{_1F_1\left(1;\frac{\gamma}{2}+1;\frac{K_+K_-}{4}\right)}\#.
	\end{equation}
	In the limit $z\longrightarrow 0$, one recovers the vacuum projector given in (\ref{vacpro000}), 
	\begin{eqnarray}
	\ket{0,\gamma}_e\,_e\langle 0,\gamma|&=&\#\frac{1}{_1F_1\left(1;\frac{\gamma}{2}+1;\frac{K_+K_-}{4}\right)}\#\nonumber
	\end{eqnarray}
	where $ pFq(a_p;b_q;0)=1$. 
	% Then, we recognize the vaccuum projector of even BGCSS.
	\item [(ii)] Odd Barut-Giradello CSs. 
	
	The odd BGCSs are characterized by the projector
	%		{\small\begin{eqnarray}
	%				\ket{z,\gamma}_o&=&\frac{1}{\sqrt{_1F_1\left(1;\frac{\gamma}{2}+1;\frac{|z|^2}{4}\right)-1}}\sum_{n=0}^{\infty}\frac{(zK_+)^{2n+1}}{4^{2n+1}\left(\frac{\gamma}{2}+1\right)_{2n+1}}\ket{0,\gamma}\cr
	%				&=&\frac{1}{\sqrt{_1F_1\left(1;\frac{\gamma}{2}+1;\frac{|z|^2}{4}\right)-1}}\#\left(_1F_1\left(1;\frac{\gamma}{2}+1;\frac{zK_+}{4}\right)-1\right)\#\ket{0,\gamma},
	%				% _o\langle z,\gamma|&=&\langle 0,\gamma|\frac{1}{\sqrt{_1F_1\left(1;\frac{\gamma}{2}+1;\frac{|z|^2}{4}\right)-1}}\#\left(_1F_1\left(1;\frac{\gamma}{2}+1;\frac{\overline{z}K_-}{4}\right)-1\right)\#
	%		\end{eqnarray}}
	% We provide 
	{\small\begin{equation}
		\ket{z,\gamma}_o\,_o\langle z,\gamma|
		% &=&\frac{\#_1F_1\left(1;\frac{\gamma}{2}+1;\frac{zK_+}{4}\right)-1\#}{\sqrt{_1F_1\left(1;\frac{\gamma}{2}+1;\frac{|z|^2}{4}\right)-1}}\ket{0,\gamma}\langle 0,\gamma|\frac{\#_1F_1\left(1;\frac{\gamma}{2}+1;\frac{\overline{z}K_-}{4}\right)-1\#}{\sqrt{_1F_1\left(1;\frac{\gamma}{2}+1;\frac{|z|^2}{4}\right)}-1}\cr
		=\frac{1}{_1F_1\left(1;\frac{\gamma}{2}+1;\frac{|z|^2}{4}\right)-1}\#\frac{\left(_1F_1\left(1;\frac{\gamma}{2}+1;\frac{\overline{z}K_-}{4}\right)-1\right)\,\left(_1F_1\left(1;\frac{\gamma}{2}+1;\frac{zK_+}{4}\right)-1\right)}{_1F_1\left(1;\frac{\gamma}{2}+1;\frac{K_+K_-}{4}\right)-1}\#.
		\end{equation}}
	In the limit $z\longrightarrow 0$, this expression reduces to the vacuum projector in (\ref{vacpro001})
	\begin{equation}
	\ket{0,\gamma}_o\,_o\langle 0,\gamma|=\#\frac{1}{_1F_1\left(1;\frac{\gamma}{2}+1;\frac{K_+K_-}{4}\right)-1}\#.\nonumber
	\end{equation}
\end{enumerate}
It is straightforward to verify that the BGCSs constructed in the Hilbert space $\mathfrak H = span\{\ket{n,\gamma}\}_{n= 0}^{\infty}$ are  eigenstates of the
lowering generator $K_-:$
{\small\begin{eqnarray}
	K_-\ket{z,\gamma}_e&=&[{\cal N}_e(|z|^2,\gamma)]^{-\frac{1}{2}}\sum_{n=0}^{\infty}\frac{z^{2n}}{\sqrt{4^{2n}\left(\frac{\gamma}{2}+1\right)_{2n}}}K_-\ket{2n,\gamma}=z\ket{z,\gamma}_e,\nonumber\\
	K_-\ket{z,\gamma}_o&=&[{\cal N}_o(|z|^2,\gamma)]^{-\frac{1}{2}}\sum_{n=0}^{\infty}\frac{z^{2n+1}}{\sqrt{4^{2n+1}\left(\frac{\gamma}{2}+1\right)_{2n+1}}}K_-\ket{2n+1,\gamma}=z\ket{z,\gamma}_o.\nonumber
	\end{eqnarray}}	

\subsubsection{Overlap of two BGCSs}

The overlaps of the two classes of BGCSs are delivered by the following expressions:
\begin{enumerate}
	\item [(i)] even BGCSs
	\begin{eqnarray}\label{overlap001}
	_e\langle z',\gamma\ket{z,\gamma}_e%&=&\langle 0,\gamma|\frac{\#_1F_1\left(1;\frac{\gamma}{2}+1;\frac{\overline{z'}K_-}{4}\right)\#}{\sqrt{_1F_1\left(1;\frac{\gamma}{2}+1;\frac{|z'|^2}{4}\right)}}\frac{\#_1F_1\left(1;\frac{\gamma}{2}+1;\frac{zK_+}{4}\right)\#}{\sqrt{_1F_1\left(1;\frac{\gamma}{2}+1;\frac{|z|^2}{4}\right)}}\ket{0,\gamma}\cr
	%				&=&\langle 0,\gamma|\frac{\#\sum_{n=0}^{\infty}\frac{(\overline{z'}zK_+K_-)^{2n}}{\left(4^{2n}\left(\frac{\gamma}{2}+1\right)_{2n}\right)^2}\#}{\sqrt{_1F_1\left(1;\frac{\gamma}{2}+1;\frac{|z'|^2}{4}\right)\,_1F_1\left(1;\frac{\gamma}{2}+1;\frac{|z|^2}{4}\right)}}\ket{0,\gamma}\cr
	%				&=&\frac{\#\sum_{n=0}^{\infty}\frac{(\overline{z'}z)^{2n}\langle 0,\gamma|(K_+K_-)^{2n}\ket{0,\gamma}}{\left(4^{2n}\left(\frac{\gamma}{2}+1\right)_{2n}\right)^2}\#}{\sqrt{_1F_1\left(1;\frac{\gamma}{2}+1;\frac{|z'|^2}{4}\right)\,_1F_1\left(1;\frac{\gamma}{2}+1;\frac{|z|^2}{4}\right)}}\cr
	%				&=&\frac{\sum_{n=0}^{\infty}\frac{(\overline{z'}z)^{2n}}{4^{2n}\left(\frac{\gamma}{2}+1\right)_{2n}}}{\sqrt{_1F_1\left(1;\frac{\gamma}{2}+1;\frac{|z'|^2}{4}\right)\,_1F_1\left(1;\frac{\gamma}{2}+1;\frac{|z|^2}{4}\right)}}\cr
	&=&\frac{_1F_1\left(1;\frac{\gamma}{2}+1;\frac{\overline{z'}z}{4}\right)}{\sqrt{_1F_1\left(1;\frac{\gamma}{2}+1;\frac{|z'|^2}{4}\right)\,_1F_1\left(1;\frac{\gamma}{2}+1;\frac{|z|^2}{4}\right)}}
	\end{eqnarray}
	\item [(ii)] odd BGCSs
	\begin{eqnarray}\label{overlap003}
	_o\langle z',\gamma\ket{z,\gamma}_o 
	%				&=&\langle 0,\gamma|\frac{\#_1F_1\left(1;\frac{\gamma}{2}+1;\frac{\overline{z'}K_-}{4}\right)-1\#}{\sqrt{_1F_1\left(1;\frac{\gamma}{2}+1;\frac{|z'|^2}{4}\right)-1}}\frac{\#_1F_1\left(1;\frac{\gamma}{2}+1;\frac{zK_+}{4}\right)-1\#}{\sqrt{_1F_1\left(1;\frac{\gamma}{2}+1;\frac{|z|^2}{4}\right)-1}}\ket{0,\gamma}\cr
	%				&=&\langle 0,\gamma|\frac{\#\sum_{n=0}^{\infty}\frac{(\overline{z'}zK_+K_-)^{2n+1}}{\left(4^{2n+1}\left(\frac{\gamma}{2}+1\right)_{2n+1}\right)^2}\#}{\sqrt{(_1F_1\left(1;\frac{\gamma}{2}+1;\frac{|z'|^2}{4}\right)-1)(\,_1F_1\left(1;\frac{\gamma}{2}+1;\frac{|z|^2}{4}\right)-1)}}\ket{0,\gamma}\cr
	%				&=&\frac{\#\sum_{n=0}^{\infty}\frac{(\overline{z'}z)^{2n+1}\langle 0,\gamma|(K_+K_-)^{2n+1}\ket{0,\gamma}}{\left(4^{2n+1}\left(\frac{\gamma}{2}+1\right)_{2n+1}\right)^2}\#}{\sqrt{(_1F_1\left(1;\frac{\gamma}{2}+1;\frac{|z'|^2}{4}\right)-1)(\,_1F_1\left(1;\frac{\gamma}{2}+1;\frac{|z|^2}{4}\right)-1)}}\cr
	%				&=&\frac{\sum_{n=0}^{\infty}\frac{(\overline{z'}z)^{2n+1}}{4^{2n+1}\left(\frac{\gamma}{2}+1\right)_{2n+1}}}{\sqrt{\left(_1F_1\left(1;\frac{\gamma}{2}+1;\frac{|z'|^2}{4}\right)-1\right)\left(\,_1F_1\left(1;\frac{\gamma}{2}+1;\frac{|z|^2}{4}\right)-1\right)}}\cr
	&=&\frac{_1F_1\left(1;\frac{\gamma}{2}+1;\frac{\overline{z'}z}{4}\right)-1}{\sqrt{\left(_1F_1\left(1;\frac{\gamma}{2}+1;\frac{|z'|^2}{4}\right)-1\right)\left(\,_1F_1\left(1;\frac{\gamma}{2}+1;\frac{|z|^2}{4}\right)-1\right)}}.
	\end{eqnarray}
\end{enumerate}	
\subsubsection{Continuity in the labeling}

The BGCSs $\ket{z,\gamma}_e$ and $\ket{z,\gamma}_o$  are continuous in labeling
$z$,  and we can show that the transformation of CSs
parameter $z' \mapsto z$ leads to the transformation of BGCSs 
$\ket{z',\gamma}_e \mapsto \ket{z,\gamma}_e$,  and $\ket{z',\gamma}_o \mapsto \ket{z,\gamma}_o$, respectively. Indeed, we have:\\
{\small
	$\mbox{If}\quad |z-z'|\longrightarrow 0
	\quad \mbox{then }$
	\begin{eqnarray}
	&&\|\ket{z,\gamma}_e - \ket{z',\gamma}_e \|^2 
	= 2\left[1-\mbox{Re}\left(_e\langle z',\gamma\ket{z,\gamma}_e\right)\right]\longrightarrow 0, \crcr
	&&\|\ket{z,\gamma}_o - \ket{z',\gamma}_o \|^2 
	= 2\left[1-\mbox{Re}\left(_o\langle z',\gamma\ket{z,\gamma}_o\right)\right]\longrightarrow 0,
	\end{eqnarray}
}	
where (\ref{overlap001}) and (\ref{overlap003}) together have been used.

\subsubsection{Resolution of the unity operator}

Next, one of the  fundamental property of any set of CSs is the resolution
of unity operator. We have the following proposition:
\bpro\label{ident000}
The constructed pair of BGCSs satisfies on the Hilbert space  $\mathfrak H = span\{\ket{n,\gamma}\}_{n= 0}^{\infty}$  the following resolutions  of the identity: 
\begin{equation}\label{ident0001}
\int_{\mathbb{C}} \frac{d^2z}{\pi}W_e(|z|^2,\gamma)\ket{z,\gamma}_e\,_e\bra{z,\gamma}=\mathbb{I_{\mathfrak H}, }\qquad
\int_{\mathbb{C}} \frac{d^2z}{\pi}W_o(|z|^2,\gamma)\ket{z,\gamma}_o\ _o\bra{z,\gamma}=\mathbb{I_{\mathfrak H}}
\end{equation}
where the appropriate weight functions  $W_e(|z|^2,\gamma)$ and $W_o(|z|^2,\gamma)$,  are obtained through the Mellin transform in the DOOT framework, and provided  as:
{\small\begin{equation}\label{weight-function00}
	W_e(|z|^2,\gamma) =
	\frac{|z|^{\gamma}e^{-\frac{|z|^2}{4}}}{2^{\gamma+2}}\MeijerG{1,1}{1,2}{0}{0;-\frac{\gamma}{2}}{-\frac{|z|^2}{4}}, 
	\end{equation}
	\begin{equation}\label{weight-function01}
	W_o(|z|^2,\gamma) = W_e(|z|^2,\gamma)-\frac{1}{4\Gamma\left(\frac{\gamma}{2}+1\right)}\MeijerG{2,0}{1,2}{-1}{-1,\frac{\gamma}{2}}{\frac{|z|^2}{4}},
	\end{equation}}
respectively.
\epro
\subsection{Gazeau-Klauder coherent states construction}
% \subsubsection{Fock basis on the DOOT framework}
We now proceed to the construction of the Gazeau-Klauder coherent states (GKCSs) associated with the isotonic oscillator within the framework of the diagonal operator ordering technique (DOOT).

Applying the DOOT prescription and using the actions of the generators $K_+$ and $K_-,$ the Fock basis vectors and their duals are expressed as
\begin{equation*}
|\Phi_n^\gamma\rangle = \frac{(K_+)^n}{\sqrt{4^n\left(\frac{\gamma}{2}+1\right)_n}}|\Phi_0^\gamma\rangle, \qquad 
\langle \Phi_n^\gamma| = \langle \Phi_0^\gamma|\frac{(K_-)^n}{\sqrt{4^n\left(\frac{\gamma}{2}+1\right)_n}},
\end{equation*}
respectively. Making use of the identity 
\begin{equation*}\label{over003}
|\Phi_n^\gamma\rangle\langle \Phi_n^\gamma|=\frac{(K_+K_-)^n}{4^n\left(\frac{\gamma}{2}+1\right)_n}|\Phi_0^\gamma\rangle\langle \Phi_0^\gamma|, 
\end{equation*}
the projector onto the vacuum Fock state can be written, within the DOOT scheme, as
\begin{equation}
% |\Phi_n^\gamma\rangle\langle \Phi_n^\gamma|&=&\frac{(K_+K_-)^n}{4^n\left(\frac{\gamma}{2}+1\right)_n}|\Phi_0^\gamma\rangle\langle \Phi_0^\gamma|\cr
% \Longrightarrow\sum_{n=0}^{\infty}|\Phi_n^\gamma\rangle\langle \Phi_n^\gamma|&=&\sum_{n=0}^{\infty}\frac{(K_+K_-)^n}{4^n\left(\frac{\gamma}{2}+1\right)_n}|\Phi_0^\gamma\rangle\langle \Phi_0^\gamma|\cr
% \mathbb{I_{\mathfrak H}}&=&\sum_{n=0}^{\infty}\frac{(K_+K_-)^n}{4^n\left(\frac{\gamma}{2}+1\right)_n}|\Phi_0^\gamma\rangle\langle \Phi_0^\gamma|\cr
% &=& _1F_1\left(1;\frac{\gamma}{2}+1;\frac{K_+K_-}{4}\right)|\Phi_0^\gamma\rangle\langle \Phi_0^\gamma|\cr
|\Phi_0^\gamma\rangle\langle \Phi_0^\gamma| = \#\frac{\mathbb{I_{\mathfrak H}}}{_1F_1\left(1;\frac{\gamma}{2}+1;\frac{K_+K_-}{4}\right)}\#.%\label{Over1}
\end{equation}
As a consequence, the expectation value $\langle \Phi_0^\gamma|(K_+K_-)^{n}|\Phi_0^\gamma\rangle$  is readily obtained in the closed form :
\begin{equation*}
\langle \Phi_0^\gamma|(K_+K_-)^{n}|\Phi_0^\gamma\rangle = 4^{n}\left(\frac{\gamma}{2}+1\right)_{n}.
\end{equation*}
The isotonic oscillator Gazeau-Klauder coherent states were originally introduced in \cite{Thirulogasanthar-Saad} and are defined as
\begin{equation}\label{gkcs1}
\ket{J,\alpha}=\frac{1}{\,_1F_1\left(1;\frac{\gamma}{2}+1;\frac{J}{4}\right)}\sum_{n=0}^{\infty}\frac{J^{n/2}}{\sqrt{4^n(\frac{\gamma}{2}+1)_n}}e^{-i e_n\alpha}\ket{\Phi_n^{\gamma}},
\end{equation}
where $J \ge 0$, $-\infty\le \alpha\le\infty$, $\left\{|\Phi_n^{\gamma}\rangle\right\}_{n = 0}^{\infty}$ denote the eigenstates of the quantum Hamiltonian, and $e_n$ are the corresponding energy eigenvalues.
Within the DOOT formalism, the GKCSs in (\ref{gkcs1}) can be cast into the compact operator form
\begin{eqnarray}
\ket{J,\alpha} 
% \frac{1}{\sqrt{\,_1F_1\left(1;\frac{\gamma}{2}+1;\frac{J}{4}\right)}}\sum_{n=0}^{\infty}\frac{J^{n/2}}{\sqrt{4^n(\frac{\gamma}{2}+1)_n}}e^{-i e_n\alpha}\ket{\Phi_n^{\gamma}}\cr 
% &=&\frac{1}{\sqrt{\,_1F_1\left(1;\frac{\gamma}{2}+1;\frac{J}{4}\right)}}\sum_{n=0}^{\infty}\frac{J^{n/2}K_+^n}{4^n(\frac{\gamma}{2}+1)_n}e^{-i (4n+2\gamma)\alpha}\ket{\Phi_0^{\gamma}}\cr 
% &=&\frac{1}{\sqrt{\,_1F_1\left(1;\frac{\gamma}{2}+1;\frac{J}{4}\right)}}\sum_{n=0}^{\infty}\frac{(e^{-4i\alpha}J^{1/2}K_+)^n}{4^n(\frac{\gamma}{2}+1)_n}e^{-2i\gamma\alpha}\ket{\Phi_0^{\gamma}}\cr 
=\frac{1}{\sqrt{\,_1F_1\left(1;\frac{\gamma}{2}+1;\frac{J}{4}\right)}}\#\,_1F_1\left(1;\frac{\gamma}{2}+1;\frac{\sqrt{J}}{4}e^{-4i\alpha}K_+\right)\#\ket{\Phi_0^{\gamma,\alpha}},\nonumber
\end{eqnarray}
where the phase-shifted vacuum state is defined by $\ket{\Phi_0^{\gamma,\alpha}}=e^{-2i\gamma\alpha}\ket{\Phi_0^{\gamma}}$.
% The GKCSs and their conjugate were :
% \begin{eqnarray}
% \ket{J,\alpha}&=&\frac{1}{\sqrt{\,_1F_1\left(1;\frac{\gamma}{2}+1;\frac{J}{4}\right)}}\#\,_1F_1\left(1;\frac{\gamma}{2}+1;\frac{\sqrt{J}}{4}e^{-4i\alpha}K_+\right)\#\ket{\Phi_0^{\gamma,\alpha}}\\
% \langle J,\alpha|&=&\frac{1}{\sqrt{\,_1F_1\left(1;\frac{\gamma}{2}+1;\frac{J}{4}\right)}}\#\,_1F_1\left(1;\frac{\gamma}{2}+1;\frac{\sqrt{J}}{4}e^{4i\alpha}K_-\right)\#\bra{\Phi_0^{\gamma,\alpha}}.
% \end{eqnarray}
The projector onto a GKCS then reads
{\small\begin{equation}
	\ket{J,\alpha}\langle J,\alpha'|
	% &=&\frac{\#\,_1F_1\left(1;\frac{\gamma}{2}+1;\frac{\sqrt{J}}{4}e^{-4i\alpha}K_+\right)\#}{\sqrt{\,_1F_1\left(1;\frac{\gamma}{2}+1;\frac{J}{4}\right)}}\ket{\Phi_0^{\gamma,\alpha}}\bra{\Phi_0^{\gamma,\alpha'}}\frac{\#\,_1F_1\left(1;\frac{\gamma}{2}+1;\frac{\sqrt{J}}{4}e^{4i\alpha'}K_-\right)\#}{\sqrt{\,_1F_1\left(1;\frac{\gamma}{2}+1;\frac{J}{4}\right)}}\cr
	% &=&\frac{\#\,_1F_1\left(1;\frac{\gamma}{2}+1;\frac{\sqrt{J}}{4}e^{-4i\alpha}K_+\right)}{\,_1F_1\left(1;\frac{\gamma}{2}+1;\frac{J}{4}\right)}e^{-2i\gamma\alpha}\ket{\Phi_0^{\gamma}}\bra{\Phi_0^{\gamma}}e^{2i\gamma\alpha'}\,_1F_1\left(1;\frac{\gamma}{2}+1;\frac{\sqrt{J}}{4}e^{4i\alpha'}K_-\right)\#\cr following properties
	=\#\frac{\,_1F_1\left(1;\frac{\gamma}{2}+1;\frac{\sqrt{J}}{4}e^{-4i\alpha}K_+\right)\,_1F_1\left(1;\frac{\gamma}{2}+1;\frac{\sqrt{J}}{4}e^{4i\alpha'}K_-\right)}{\,_1F_1\left(1;\frac{\gamma}{2}+1;\frac{J}{4}\right)\,_1F_1\left(1;\frac{\gamma}{2}+1;\frac{K_+K_-}{4}\right)}e^{-2i\gamma(\alpha-\alpha')}\#.\label{Opgkcs}
	\end{equation}}
By setting $J = 0$ in (\ref{Opgkcs}),one immediately recovers the projector onto the vacuum state,
\begin{equation*}
\ket{0,\alpha}\langle 0,\alpha'|
% &=&\#\frac{\,_1F_1\left(1;\frac{\gamma}{2}+1;\frac{\sqrt{0 }}{4}e^{-4i\alpha}K_+\right)\,_1F_1\left(1;\frac{\gamma}{2}+1;\frac{\sqrt{0}}{4}e^{4i\alpha'}K_-\right)}{\,_1F_1\left(1;\frac{\gamma}{2}+1;\frac{0}{4}\right)\,_1F_1\left(1;\frac{\gamma}{2}+1;\frac{K_+K_-}{4}\right)}e^{-2i\gamma(\alpha-\alpha')}\#\cr
% &=&
= \#\frac{e^{-2i\gamma(\alpha-\alpha')}}{\,_1F_1\left(1;\frac{\gamma}{2}+1;\frac{K_+K_-}{4}\right)}\#.\label{vacgkcs}
\end{equation*}
Furthermore, in the limit $\alpha\longrightarrow\alpha'$, (\ref{Opgkcs}) reduces to
{\small\begin{eqnarray}
	\ket{J,\alpha}\langle J,\alpha|&=&\#\frac{\,_1F_1\left(1;\frac{\gamma}{2}+1;\frac{\sqrt{J}}{4}e^{-4i\alpha}K_+\right)\,_1F_1\left(1;\frac{\gamma}{2}+1;\frac{\sqrt{J}}{4}e^{4i\alpha}K_-\right)}{\,_1F_1\left(1;\frac{\gamma}{2}+1;\frac{J}{4}\right)\,_1F_1\left(1;\frac{\gamma}{2}+1;\frac{K_+K_-}{4}\right)}\#.
	\end{eqnarray}}

\subsubsection*{Overlap of two GKCSs}
The scalar product of a ket and its bra counterpart leads to
\begin{equation}\label{scal000}
\langle J',\alpha'\ket{J,\alpha}
% &=&\langle \Phi_0^\gamma|\frac{\#_1F_1\left(1;\frac{\gamma}{2}+1;\frac{J'K_-e^{4i\alpha'}}{4}\right)\#}{\sqrt{_1F_1\left(1;\frac{\gamma}{2}+1;\frac{J'}{4}\right)}}\frac{\#_1F_1\left(1;\frac{\gamma}{2}+1;\frac{JK_+e^{-4i\alpha}}{4}\right)\#e^{-2i\gamma(\alpha-\alpha')}}{\sqrt{_1F_1\left(1;\frac{\gamma}{2}+1;\frac{J}{4}\right)}}\ket{\Phi_0^\gamma}\cr
% &=&\langle \Phi_0^\gamma|\frac{\#\sum_{n=0}^{\infty}\frac{(\sqrt{J'J}K_-K_+e^{-4i(\alpha-\alpha')})^{n}}{\left(4^{n}\left(\frac{\gamma}{2}+1\right)_{n}\right)^2}\#}{\sqrt{_1F_1\left(1;\frac{\gamma}{2}+1;\frac{J'}{4}\right)\,_1F_1\left(1;\frac{\gamma}{2}+1;\frac{J}{4}\right)}}\ket{\Phi_0^\gamma}\cr
% &=&\frac{\#\sum_{n=0}^{\infty}\frac{(\sqrt{J'J}e^{-4i(\alpha-\alpha')})^n\langle \Phi_0^\gamma|(K_-K_+)^{n}\ket{\Phi_0^\gamma}}{\left(4^{n}\left(\frac{\gamma}{2}+1\right)_{n}\right)^2}\#e^{-2i\gamma(\alpha-\alpha')}}{\sqrt{_1F_1\left(1;\frac{\gamma}{2}+1;\frac{J'}{4}\right)\,_1F_1\left(1;\frac{\gamma}{2}+1;\frac{J}{4}\right)}}\cr
% &=&\frac{\sum_{n=0}^{\infty}\frac{(\sqrt{J'J}e^{-4i(\alpha-\alpha')})^{n}}{4^{n}\left(\frac{\gamma}{2}+1\right)_{n}}e^{-2i\gamma(\alpha-\alpha')}}{\sqrt{_1F_1\left(1;\frac{\gamma}{2}+1;\frac{J'}{4}\right)\,_1F_1\left(1;\frac{\gamma}{2}+1;\frac{J}{4}\right)}}\cr
= \frac{_1F_1\left(1;\frac{\gamma}{2}+1;\frac{\sqrt{J'J}}{4}e^{4i(\alpha'-\alpha)}\right)e^{2i\gamma(\alpha'-\alpha)}}{\sqrt{_1F_1\left(1;\frac{\gamma}{2}+1;\frac{J'}{4}\right)\,_1F_1\left(1;\frac{\gamma}{2}+1;\frac{J}{4}\right)}},
\end{equation}
such that taking  $\alpha\longrightarrow\alpha'$ provides
\begin{equation}
\langle J',\alpha\ket{J,\alpha}=\frac{_1F_1\left(1;\frac{\gamma}{2}+1;\frac{\sqrt{J'J}}{4}\right)}{\sqrt{_1F_1\left(1;\frac{\gamma}{2}+1;\frac{J'}{4}\right)\,_1F_1\left(1;\frac{\gamma}{2}+1;\frac{J}{4}\right)}}.
\end{equation}	
% \section{Some mathematical properties}
% \subsection{Overlap}
%\subsubsection*{Resolution of the unity operator}
We have the following property:
\bpro\label{ident001}
The constructed GKCSs satisfy on the Hilbert space  $\mathfrak H = span\{\ket{\Phi}_n^\gamma\}_{n= 0}^{\infty}$  the following resolutions  of the identity: 
\begin{equation}\label{ident003}
\int_{\mathbb{C}}{\cal N}(J)^{2}\lambda(J)\ket{J,\alpha}\bra{J,\alpha}dJ d\alpha=\mathbb{I_{\mathfrak H}}, \quad {\cal N}(J)^{2} = \,_1F_1\left(1;\frac{\gamma}{2}+1;\frac{J}{4}\right), 
\end{equation}
where the appropriate weight function,   obtained through the Mellin transform in the DOOT framework is : $\lambda(J)=\frac{1}{4\Gamma\left(\frac{\gamma}{2}+1\right)}\MeijerG{2,0}{1,2}{-1}{-1;\frac{\gamma}{2}}{\frac{|z|^2}{4}}$.

\epro

\begin{figure}[h]
	\begin{subfigure}[b]{0.3\textwidth}
		\includegraphics[width=\textwidth]{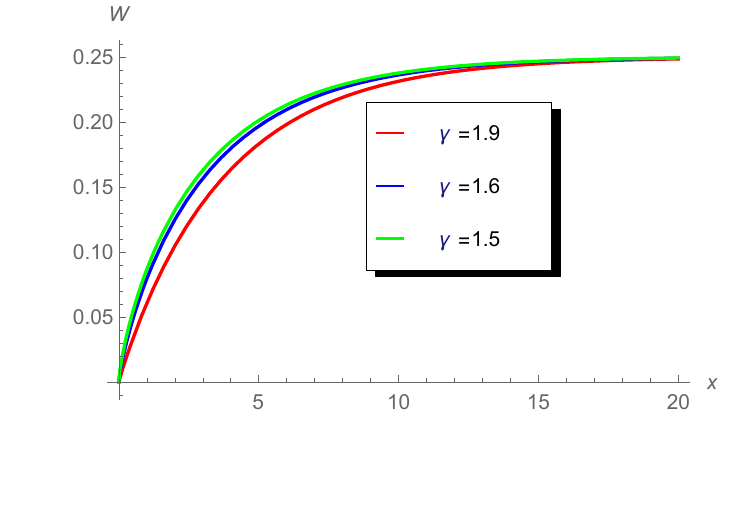}
		\caption{Even coherent states}
		%\label{fig:gull}
	\end{subfigure}
	\begin{subfigure}[b]{0.3\textwidth}
		\includegraphics[width=\textwidth]{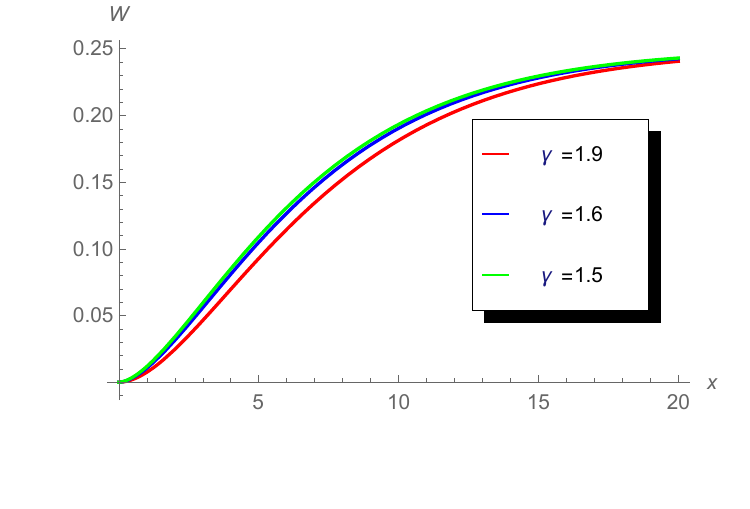}
		\caption{Odd coherent states}
		%	\label{fig:gull}
	\end{subfigure}
	\begin{subfigure}[b]{0.3\textwidth}
		\includegraphics[width=\textwidth]{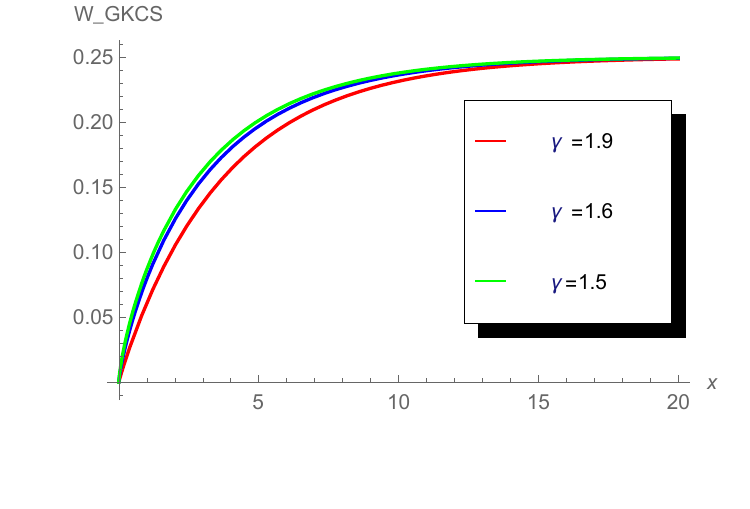}
		\caption{GK coherent states}
		%	\label{fig:gull}
	\end{subfigure}
	\caption{\it \small  
		The plot of the weight  functions $W_e(|z|^2,\gamma)$ \textnormal{(a)}, $W_o(|z|^2,\gamma)$ \textnormal{(b)}, 
		and $\lambda(J)$  \textnormal{(c)} against $x = |z|$,  for different  values of the Bargmann index  $\gamma$.}
\end{figure} 
The weight functions $W_e(|z|^2,\gamma)$  (\ref{weight-function00}) and $W_o(|z|^2,\gamma)$  \ref{weight-function01}), are always increasing functions of $x = |z|$.  The curves are positive, as displayed  by the graphs (a) and (b) in  Figure 1, and confirm the positivity of the weight functions for any value of  the Bargmann index $\gamma > 0$, which increases the asymptotic behaviour of the curves   by taking smaller values.

In the next paragraph, we  will examine some analytic properties of the BGCSs and GKCSs in the light of the DOOT rules.

\section{Reproducing kernel}

The Fock Hilbert space $\mathfrak H = span\{\ket{\Phi}_n^\gamma\}_{n= 0}^{\infty}$  is also a reproducing kernel Hilbert space with
reproducing kernel. The corresponding reproducing kernels for each set of CSs are provided as: 
\begin{enumerate}
	\item For even BGCSs
	\begin{equation}\label{kern000}
	K_e(z',z)
	% 	&=& _e\langle z',\gamma|z,\gamma\rangle_e\cr
	% 	&=&\langle 0,\gamma|\frac{\#_1F_1\left(1;\frac{\gamma}{2}+1;\frac{\overline{z'}K_-}{4}\right)\#}{\sqrt{_1F_1\left(1;\frac{\gamma}{2}+1;\frac{|z|^2}{4}\right)}}\frac{\#_1F_1\left(1;\frac{\gamma}{2}+1;\frac{zK_+}{4}\right)\#}{\sqrt{_1F_1\left(1;\frac{\gamma}{2}+1;\frac{|z'|^2}{4}\right)}}\ket{0,\gamma}\cr
	% 	&=&\langle 0,\gamma|\#\frac{\sum_{n=0}^{\infty}\frac{(zK_+)^{2n}(\overline{z'}K_-)^{2n}}{(4^{2n})^2\left(\frac{\gamma}{2}+1\right)_{2n}^2}}{\sqrt{_1F_1\left(1;\frac{\gamma}{2}+1;\frac{|z|^2}{4}\right)\,_1F_1\left(1;\frac{\gamma}{2}+1;\frac{|z'|^2}{4}\right)}}\#\ket{0,\gamma}\cr
	% 	&=&\frac{\sum_{n=0}^{\infty}\frac{(z\overline{z'})^{2n}}{(4^{2n})^2\left(\frac{\gamma}{2}+1\right)_{2n}^2}\langle 0,\gamma|\#(K_+K_-)^{2n}\#\ket{0,\gamma}}{\sqrt{_1F_1\left(1;\frac{\gamma}{2}+1;\frac{|z|^2}{4}\right)\,_1F_1\left(1;\frac{\gamma}{2}+1;\frac{|z'|^2}{4}\right)}}\cr
	% 	&=&\frac{\sum_{n=0}^{\infty}\frac{(z\overline{z'})^{2n}}{(4^{2n})^2\left(\frac{\gamma}{2}+1\right)_{2n}^2}4^{2n}\left(\frac{\gamma}{2}+1\right)_{2n}}{\sqrt{_1F_1\left(1;\frac{\gamma}{2}+1;\frac{|z|^2}{4}\right)\,_1F_1\left(1;\frac{\gamma}{2}+1;\frac{|z'|^2}{4}\right)}}\cr
	% 	&=&\frac{\sum_{n=0}^{\infty}\frac{(z\overline{z'})^{2n}}{4^{2n}\left(\frac{\gamma}{2}+1\right)_{2n}}}{\sqrt{_1F_1\left(1;\frac{\gamma}{2}+1;\frac{|z|^2}{4}\right)\,_1F_1\left(1;\frac{\gamma}{2}+1;\frac{|z'|^2}{4}\right)}}\cr
	=
	\frac{_1F_1\left(1;\frac{\gamma}{2}+1;\frac{z\overline{z'}}{4}\right)}{\sqrt{_1F_1\left(1;\frac{\gamma}{2}+1;\frac{|z|^2}{4}\right)\,_1F_1\left(1;\frac{\gamma}{2}+1;\frac{|z'|^2}{4}\right)}};
	\end{equation}
	\item For odd BGCSs
	\begin{equation}\label{kern001}
	K_o(z',z)
	% 	&=&_o\langle z',\gamma|z,\gamma\rangle_o\cr
	% 	&=&\langle 0,\gamma|\frac{\#_1F_1\left(1;\frac{\gamma}{2}+1;\frac{\overline{z'}K_-}{4}\right)-1\#}{\sqrt{_1F_1\left(1;\frac{\gamma}{2}+1;\frac{|z|^2}{4}\right)-1}}\frac{\#_1F_1\left(1;\frac{\gamma}{2}+1;\frac{zK_+}{4}\right)-1\#}{\sqrt{_1F_1\left(1;\frac{\gamma}{2}+1;\frac{|z'|^2}{4}\right)-1}}\ket{0,\gamma}\cr
	% 	&=&\langle 0,\gamma|\#\frac{\sum_{n=0}^{\infty}\frac{(zK_+)^{2n+1}(\overline{z'}K_-)^{2n+1}}{(4^{2n+1})^2\left(\frac{\gamma}{2}+1\right)_{2n+1}^2}}{\sqrt{\left[_1F_1\left(1;\frac{\gamma}{2}+1;\frac{|z|^2}{4}\right)-1\right]\left[\,_1F_1\left(1;\frac{\gamma}{2}+1;\frac{|z'|^2}{4}\right)-1\right]}}\#\ket{0,\gamma}\cr
	% 	&=&\frac{\sum_{n=0}^{\infty}\frac{(z\overline{z'})^{2n+1}}{(4^{2n+1})^2\left(\frac{\gamma}{2}+1\right)_{2n+1}^2}\langle 0,\gamma|\#(K_+K_-)^{2n}\#\ket{0,\gamma}}{\sqrt{\left[_1F_1\left(1;\frac{\gamma}{2}+1;\frac{|z|^2}{4}\right)-1\right]\left[\,_1F_1\left(1;\frac{\gamma}{2}+1;\frac{|z'|^2}{4}\right)-1\right]}}\cr
	% 	&=&\frac{\sum_{n=0}^{\infty}\frac{(z\overline{z'})^{2n+1}}{(4^{2n+1})^2\left(\frac{\gamma}{2}+1\right)_{2n+1}^2}4^{2n+1}\left(\frac{\gamma}{2}+1\right)_{2n+1}}{\sqrt{\left[_1F_1\left(1;\frac{\gamma}{2}+1;\frac{|z|^2}{4}\right)-1\right]\left[\,_1F_1\left(1;\frac{\gamma}{2}+1;\frac{|z'|^2}{4}\right)-1\right]}}\cr
	% 	&=&\frac{\sum_{n=0}^{\infty}\frac{(z\overline{z'})^{2n+1}}{4^{2n+1}\left(\frac{\gamma}{2}+1\right)_{2n+1}}}{\sqrt{\left[_1F_1\left(1;\frac{\gamma}{2}+1;\frac{|z|^2}{4}\right)-1\right]\left[\,_1F_1\left(1;\frac{\gamma}{2}+1;\frac{|z'|^2}{4}\right)-1\right]}}\cr
	% 	
	= \frac{_1F_1\left(1;\frac{\gamma}{2}+1;\frac{z\overline{z'}}{4}\right)-1}{\sqrt{\left[_1F_1\left(1;\frac{\gamma}{2}+1;\frac{|z|^2}{4}\right)-1\right]\left[\,_1F_1\left(1;\frac{\gamma}{2}+1;\frac{|z'|^2}{4}\right)-1\right]}};
	\end{equation}
	\item For GKCSs
	\begin{equation}\label{kern003}
	K_{GKCS}(J',\alpha',J,\alpha) 
	%  &=&\langle J,\alpha| J',\alpha'\rangle\cr &=&\bra{\Phi_0^{\gamma,\alpha'}}\frac{\#\,_1F_1\left(1;\frac{\gamma}{2}+1;\frac{\sqrt{J'}}{4}e^{-4i\alpha'}K_+\right)}{\sqrt{\,_1F_1\left(1;\frac{\gamma}{2}+1;\frac{J'}{4}\right)}}\frac{\,_1F_1\left(1;\frac{\gamma}{2}+1;\frac{\sqrt{J}}{4}e^{4i\alpha'}K_-\right)\#}{\sqrt{\,_1F_1\left(1;\frac{\gamma}{2}+1;\frac{J}{4}\right)}}\ket{\Phi_0^{\gamma,\alpha}}\cr
	%  
	=\frac{\,_1F_1\left(1;\frac{\gamma}{2}+1;\frac{\sqrt{J'J}}{4}\right)}{\sqrt{\,_1F_1\left(1;\frac{\gamma}{2}+1;\frac{J'}{4}\right)\,_1F_1\left(1;\frac{\gamma}{2}+1;\frac{J}{4}\right)}}e^{-4i(\alpha'-\alpha)}.
	\end{equation}
\end{enumerate}

Whenever we have an orthonormal basis spanning the Fock Hilbert space $\mathfrak{H}$, from  (\ref{kern000})-(\ref{kern003}), and  the overcompleteness identites (\ref{ident0001}) and (\ref{ident003}) satisfied by the BGCSs and GKCSs, we have:
\bpro\label{kernel000}
The following properties
\begin{enumerate}
	\item [(i)] hermiticity $\overline{ K_e(z,z')} =  K_e(z',z)$,  \qquad $\overline{ K_o(z,z')} =  K_o(z',z),$  \qquad $\overline{ K_{GKCS}(z,z')} =  K_o(z',z),$
	\item  [(ii)] positivity $  K_e(z,z) > 0$,  \qquad $K_o(z,z) > 0,$\qquad $K_{GKCS}(z,z) > 0,$
	\item  [(iii)] idempotence  
	{\small
		\begin{eqnarray}
		&& \int_{\mathbb{C}} K_e(z,z") K_e(z",z')\frac{W_e(|z"|^2)d^2z"}{\pi} = K_e(z,z')\nonumber, \cr 
		&&\int_{\mathbb{C}} K_o(z,z") K_o(z",z')\frac{W_o(|z"|^2)d^2z"}{\pi} =  K_o(z,z'),\\
		&&\int_{\mathbb{C}} K_{GKCS}(J,J") K_{GKCS}(J",J')\frac{W_{GKCS}(|J"|^2)dJ"d\alpha}{\pi} =  K_{GKCS}(J,J')\nonumber
		\end{eqnarray}
	}
\end{enumerate}
are satisfied by $\mathcal K_e$ and $\mathcal K_o$ on the Hilbert space $\mathfrak H$.
\epro

\section{Expectations values}
The constructed BGCSs and GKCSs can be used in different physical applications to calculate
the expectation (mean) values of any significant physical observable $\cal O$ which
characterizes the quantum system. Let us perform the expectation values of the different generators by following the DOOT rules. We get the following expressions:
\begin{enumerate}
	\item [(a)]	For the Barut-Girardello CSs
	\begin{enumerate}
		\item [(i)] Even CSs
		
		The generators $K_+$ and $K_-$ of $SU(1,1)$ act in the Fock basis as follows
		\begin{equation*}
		K_+|2n,\gamma\rangle=\sqrt{4n(2n+\gamma)}|2n+2,\gamma\rangle,\qquad
		K_-|2n,\gamma\rangle=\sqrt{4n(2n+\gamma-2)}|2n-2,\gamma\rangle.
		\end{equation*}
		We get
		\begin{equation*}
		K_-|z,\gamma\rangle_e=z|z,\gamma\rangle\qquad
		_e\langle z,\gamma|K_+=\langle z,\gamma|\overline{z}
		\end{equation*}
		%than, we obtain
		\begin{equation*}
		_e\langle z,\gamma|K_+ K_-|z,\gamma\rangle_e=|z|^2\qquad\text{and}\qquad_e\langle z,\gamma|F(K_+ K_-)|z,\gamma\rangle_e=F(|z|^2),
		\end{equation*}	
		and the following expression of the normalization function 		
		\begin{eqnarray}
		|_e\langle z,\gamma|0,\gamma\rangle_e|^2%%&=&|_e\langle z,\gamma|0,\gamma\rangle\langle 0, \gamma|z,\gamma\rangle_e|\cr
		%	&=&|_e\langle z,\gamma|\#\frac{1}{\,_1F_1\left(1;\frac{\gamma}{2}+1;\frac{K_+K_-}{4}\right)}\#|z,\gamma\rangle_e|\cr 
		&=&\frac{1}{\,_1F_1\left(1;\frac{\gamma}{2}+1;\frac{|z|^2}{4}\right)}\nonumber
		\end{eqnarray}
		\item [(ii)] Odd CSs
		
		The action of the operators in the Fock basis is as follows
		\begin{eqnarray}
		K_+|2n+1,\gamma\rangle&=&\sqrt{2(2n+1)(2n+\gamma+1)}|2n+3,\gamma\rangle\cr
		K_-|2n+1,\gamma\rangle&=&\sqrt{2(2n+1)(2n+\gamma-1)}|2n-1,\gamma\rangle\nonumber
		\end{eqnarray}
		and  on the odd CSs:
		\begin{equation*}
		K_-|z,\gamma\rangle_o=z|z,\gamma\rangle_o\qquad
		_o\langle z,\gamma|K_+=_o\langle z,\gamma|\overline{z}.
		\end{equation*}
		We get the following expressions of the mean values 
		\begin{eqnarray}
		_o\langle z,\gamma|K_+ K_-|z,\gamma\rangle_o=|z|^2\qquad
		_o\langle z,\gamma|F(K_+ K_-)|z,\gamma\rangle_o=F(|z|^2)
		\end{eqnarray}
		and the normmalization function
		\begin{eqnarray}
		|_o\langle z,\gamma|0,\gamma\rangle_o|^2%%&=&|_o\langle z,\gamma|0,\gamma\rangle\langle 0, \gamma|z,\gamma\rangle_o|\cr
		%ù&=&|_o\langle z,\gamma|\#\frac{1}{\,_1F_1\left(1;\frac{\gamma}{2}+1;\frac{K_+K_-}{4}\right)-1}\#|z,\gamma\rangle_o|\cr 
		&=&\frac{1}{\,_1F_1\left(1;\frac{\gamma}{2}+1;\frac{|z|^2}{4}\right)-1}
		\end{eqnarray}
	\end{enumerate}
	\item [(b)]{Case of the GKCSs}
	
	The expressions previously obtained for the BGCSs become as follows in the case of GKCSs
	\begin{eqnarray}
	K_+|\Phi_n^{\gamma}\rangle=\sqrt{2(n+1)(n+\gamma)}|\Phi_{n+1}^{\gamma}\rangle\qquad
	K_-|\Phi_n^{\gamma}\rangle=\sqrt{2n(n+\gamma-1)}|\Phi_{n-1}^{\gamma}\rangle\nonumber
	\end{eqnarray}
	and 
	\begin{eqnarray}
	K_-|J,\alpha\rangle=J^{\frac{1}{2}}e^{-i\alpha}|J,\alpha\rangle\qquad
	\langle J,\alpha'|K_+=J^{\frac{1}{2}}e^{i\alpha'}\langle J,\alpha'|.\nonumber
	\end{eqnarray}
	The expectation values of the operators are provided in the following relations
	\begin{eqnarray}
	\langle J,\alpha'|K_+K_-|J,\alpha\rangle=Je^{-i(\alpha-\alpha')}
	\qquad \langle J,\alpha'|F(K_+ K_-)|J,\alpha\rangle=F(Je^{-i(\alpha-\alpha')}).\nonumber
	\end{eqnarray}
	When $\alpha=\alpha'$, we obtain
	\begin{eqnarray}
	\langle J,\alpha|K_+ K_-|J,\alpha\rangle=J\qquad
	\langle J,\alpha|F(K_+ K_-)|J,\alpha\rangle=F(J)\nonumber
	\end{eqnarray}
	and the normalization function becomes
	\begin{eqnarray}
	|\langle J,\alpha|\Phi_0^{\gamma}\rangle|^2%%%&=&|\langle J,\alpha|\Phi_0^{\gamma}\rangle\langle\Phi_0^{\gamma}|J,\alpha\rangle|\cr
	%%%	&=&|\langle J,\alpha|\#\frac{1}{\,_1F_1\left(1;\frac{\gamma}{2}+1;\frac{K_+K_-}{4}\right)}\#|J,\alpha\rangle|\cr 
	&=&\frac{1}{\,_1F_1\left(1;\frac{\gamma}{2}+1;\frac{J}{4}\right)}.
	\end{eqnarray}
\end{enumerate}
\section{Photon number distribution and density of probability}
This paragraph explores the photon number distribution (PND) and the density temporel probability of the  constructed BGCSs. We analyse how these states do evolve in time under the action of the time evolution.
\subsection{Photon number distribution}
We calculate the photon number distribution which provides the probability of a field being in the constructed BGCSs ant GKCSs to contain $n$ photons in the framework of DOOT. Its oscillatory behavior is a nonclassical property of the states.  We get the following expressions :
\begin{enumerate}
	\item [(1)] For the Barut-Girardello CSs (BGCSs)
	%%\begin{enumerate}
	%	\item [(a)] Even CSs
	\begin{eqnarray}
	P_n\left( |z, \gamma\rangle_e \right) %%&=& \left|_e \langle 2n, \gamma | z, \gamma \rangle_e \right|^2 \cr
	%&=& \left| _e\langle z, \gamma | 2n, \gamma \rangle \langle 2n, \gamma | z, \gamma \rangle_e \right| \cr
	%	&=& \left| _e\langle z, \gamma | \frac{(K_-K_+)^{2n}}{4^{2n} \left( \frac{\gamma}{2} + 1 \right)_{2n}} | 0, \gamma \rangle \langle 0, \gamma | z, \gamma \rangle_e \right| \cr
	%	&=&\left| _e\langle z, \gamma |\frac{(K_-K_+)^{2n}}{4^{2n} \left( \frac{\gamma}{2} + 1 \right)_{2n}}\frac{1}{{}_1F_1\left(1; \frac{\gamma}{2} + 1; \frac{K_-K_+}{4} \right)}| z, \gamma \rangle_e \right| \cr
	&=& \frac{|z|^{4n}}{4^{2n} \left( \frac{\gamma}{2} + 1 \right)_{2n}} \cdot \frac{1}{{}_1F_1\left(1; \frac{\gamma}{2} + 1; \frac{|z|^2}{4} \right)}\\
	%	\end{eqnarray}
	%\item [(b)] Odd CSs
	%%\begin{eqnarray}
	P_n\left( |z, \gamma\rangle_o \right) %%&=& \left| _o\langle 2n+1, \gamma | z, \gamma \rangle_o \right|^2 \cr
	%&=& \left|_o\langle z, \gamma | 2n+1, \gamma \rangle \langle 2n+1, \gamma | z, \gamma \rangle_o\right| \cr
	%&=& \left| _o\langle z, \gamma | \frac{(K_-K_+)^{2n+1}}{4^{2n+1} \left( \frac{\gamma}{2} + 1 \right)_{2n+1}} | 0, \gamma \rangle \langle 0, \gamma | z, \gamma \rangle_o \right| \cr
	%%&=&\left| _o\langle z, \gamma |\frac{(K_-K_+)^{2n+1}}{4^{2n+1} \left( \frac{\gamma}{2} + 1 \right)_{2n+1}}\frac{1}{{}_1F_1\left(1; \frac{\gamma}{2} + 1; \frac{K_-K_+}{4} \right)-1}| z, \gamma \rangle_o \right| \cr
	&=& \frac{|z|^{4n+1}}{4^{2n+1} \left( \frac{\gamma}{2} + 1 \right)_{2n+1}} \cdot \frac{1}{{}_1F_1\left(1; \frac{\gamma}{2} + 1; \frac{|z|^2}{4} \right) - 1}.
	\end{eqnarray}
	%%\end{enumerate}
	\item [(2)] For the Gazeau Klauder CSs (GKCSs)
	\begin{eqnarray}
	P_n\left( |J, \alpha\rangle \right) %%&=& \left| \langle \Phi_n^\gamma | J, \alpha \rangle \right|^2 \cr
	%							&=& \left| \langle J, \alpha | \Phi_n^\gamma \rangle \langle \Phi_n^\gamma | J, \alpha \rangle \right| \cr
	%							&=&\left|\langle J, \alpha |\frac{(K_-K_+)^{n/2}}{4^{n} \left( \frac{\gamma}{2} + 1 \right)_{n}}\frac{1}{{}_1F_1\left(1; \frac{\gamma}{2} + 1; \frac{K_-K_+}{4} \right)}| J, \alpha \rangle \right| \cr
	&=& \frac{J^n}{4^n \left( \frac{\gamma}{2} + 1 \right)_n} \cdot \frac{1}{{}_1F_1\left(1; \frac{\gamma}{2} + 1; \frac{J}{4} \right)}.
	\end{eqnarray}
\end{enumerate}
\subsection{Density probability}
In this paragraph, we analyze the time evolution of these states under the action of the time evolution operator generated by the physical Hamiltonian of the quantum system. The corresponding time-dependent behavior is given by
\begin{equation}
g_{z_0}(z, t) = \left| \langle z, \gamma | e^{-iHt} | z_0, \gamma \rangle \right|^2.
\end{equation}
By acting the evolution operator $U(t)=e^{-iHt}$  on the state $| z_0, \gamma \rangle$,  we obtain
\begin{eqnarray}
| z_o, \gamma,t \rangle%&=&e^{-iHt} | z_0, \gamma \rangle\cr
%						&=&e^{-iHt}\#\frac{_1F_1\left(1; \frac{\gamma}{2} + 1; \frac{zK_+}{4} \right)}{\sqrt{_1F_1\left(1; \frac{\gamma}{2} + 1; \frac{|z|^2}{4} \right)}}\#|0,\gamma\rangle\cr
&=&e^{-2iK_0t}\#\frac{_1F_1\left(1; \frac{\gamma}{2} + 1; \frac{zK_+}{4} \right)}{\sqrt{_1F_1\left(1; \frac{\gamma}{2} + 1; \frac{|z|^2}{4} \right)}}\#|0,\gamma\rangle.\label{density002}
\end{eqnarray}
From expression (\ref{density002}), we can compute the probability density for these temporally coherent states. We obtain the following results :
\begin{enumerate}
	\item [(1)] for Barut-Girardello CSs
	%	\begin{enumerate}
	%	\item [(a)] Even CSs
	\begin{eqnarray}								g^e_{z_0}(z, t)% &=& \left| \langle z, \gamma | e^{-iHt} | z_0, \gamma \rangle_e \right|^2 \\
	%								&=& \left| \langle z_0, \gamma | e^{iHt} | z, \gamma \rangle_e \cdot _e\langle z, \gamma | e^{-iHt} | z_0, \gamma \rangle_e \right| \cr
	%								&=&\left|_e\langle z_0,\gamma|\#e^{-2iK_0t}\frac{_1F_1\left(1; \frac{\gamma}{2} + 1; \frac{\overline{z}K_-}{4}\right)}{\sqrt{_1F_1\left(1; \frac{\gamma}{2} + 1; \frac{|z|^2}{4} \right)}}\right.\cr
	%								&&\left.|0,\gamma\rangle \langle 0,\gamma| \frac{_1F_1\left(1; \frac{\gamma}{2} + 1; \frac{zK_+}{4} \right)}{\sqrt{_1F_1\left(1; \frac{\gamma}{2} + 1; \frac{|z|^2}{4} \right)}} e^{2iK_0t}\#| z_0, \gamma \rangle_e \right| \cr
	&=& \frac{{}_1F_1\left(1; \frac{\gamma}{2}+1; \frac{\overline{z} z_0(t)}{4} \right) \cdot {}_1F_1\left(1; \frac{\gamma}{2}+1; \frac{z \overline{z_0(t)}}{4} \right)}{{}_1F_1\left(1; \frac{\gamma}{2}+1; \frac{|z|^2}{4} \right) \cdot {}_1F_1\left(1; \frac{\gamma}{2}+1; \frac{|z_0(t)|^2}{4} \right)}\\
	g^o_{z_0}(z, t) 
	&=& \frac{\left[ {}_1F_1\left(1; \frac{\gamma}{2}+1; \frac{\overline{z} z_0(t)}{4} \right) - 1 \right] \cdot \left[ {}_1F_1\left(1; \frac{\gamma}{2}+1; \frac{z \overline{z_0(t)}}{4} \right) - 1 \right]}{\left[ {}_1F_1\left(1; \frac{\gamma}{2}+1; \frac{|z|^2}{4} \right) - 1 \right] \cdot \left[ {}_1F_1\left(1; \frac{\gamma}{2}+1; \frac{|z_0(t)|^2}{4} \right) - 1 \right]}
	\end{eqnarray}
	where : $z_0(t) = z_0 e^{-2it}$,
	%	\item [(b)] Odd CSs
	%%	\begin{eqnarray}
	%								g^o_{z_0}(z, t) &=& \left|_o\langle z, \gamma | e^{-iHt} | z_0, \gamma \rangle_o \right|^2 \cr
	%								&=& \left|_o\langle z_0, \gamma | e^{iHt} | z, \gamma \rangle_o \cdot _o\langle z, \gamma | e^{-iHt} | z_0, \gamma \rangle_o \right| \cr
	%								&=&\left|_o\langle z_0,\gamma|\#e^{-2iK_0t}\frac{_1F_1\left(1; \frac{\gamma}{2} + 1; \frac{\overline{z}K_-}{4}\right)-1}{\sqrt{_1F_1\left(1; \frac{\gamma}{2} + 1; \frac{|z|^2}{4} \right)-1}}|0,\gamma\rangle \langle 0,\gamma|\right.\cr
	%								&&\left.\frac{_1F_1\left(1; \frac{\gamma}{2} + 1; \frac{zK_+}{4} \right)-1}{\sqrt{_1F_1\left(1; \frac{\gamma}{2} + 1; \frac{|z|^2}{4} \right)-1}} e^{2iK_0t}\#| z_0, \gamma \rangle_o \right| \cr
	%								&=& \frac{\left[ {}_1F_1\left(1; \frac{\gamma}{2}+1; \frac{\overline{z} z_0(t)}{4} \right) - 1 \right] \cdot \left[ {}_1F_1\left(1; \frac{\gamma}{2}+1; \frac{z \overline{z_0(t)}}{4} \right) - 1 \right]}{\left[ {}_1F_1\left(1; \frac{\gamma}{2}+1; \frac{|z|^2}{4} \right) - 1 \right] \cdot \left[ {}_1F_1\left(1; \frac{\gamma}{2}+1; \frac{|z_0(t)|^2}{4} \right) - 1 \right]}
	%%	\end{eqnarray}
	%%%	\end{enumerate}
	\item [(2)]for Gazeau Klauder CSs
	\begin{eqnarray}
	g^{GK}_{z_0}(z, t) %&=& \left| \langle J, \alpha | e^{-iHt} | J_0, \alpha \rangle \right|^2 \\
	%	&=& \left| \langle J_0, \alpha | e^{iHt} | J, \alpha \rangle \cdot \langle J, \alpha | e^{-iHt} | J_0, \alpha \rangle \right| \\
	&=& \frac{{}_1F_1\left(1; \frac{\gamma}{2}+1; \frac{\sqrt{J J_0} e^{4it}}{4} \right) \cdot {}_1F_1\left(1; \frac{\gamma}{2}+1; \frac{\sqrt{J J_0} e^{-4it}}{4} \right)}{{}_1F_1\left(1; \frac{\gamma}{2}+1; \frac{J}{4} \right) \cdot {}_1F_1\left(1; \frac{\gamma}{2}+1; \frac{J_0}{4} \right)}.
	\end{eqnarray}
\end{enumerate}
\section{Quantization of elementary classical observables}
This paragraph is devoted to the quantization procedure carried out in the complex plane. As shown in Section 3, the constructed families of coherent states (CSs) provide a resolution of the identity. As an immediate consequence, we establish in this section the correspondence (quantization) between classical and quantum observables. For further details on the quantization procedure, see \cite{Gazeau, aremua-gazeau-hk, gazeauetal} and the references cited therein.

The general procedure consists of considering  $(X, \mu)$ as a measure space and the Hilbert space $L^2(X, \mu)$ defined by
\begin{equation}\label{set000}
\left\{ f : X \rightarrow \mathbb{C} | \int_{X} |f(x)|^2 d\mu(x) < \infty \right\}.
\end{equation}
The Berezin-Toeplitz or anti-Wick or coherent state quantization, as used by various authors in	the literature, associates a classical observable given by a function  $f(x) \in L^2(X, \mu)$  on $X$ to an operator valued
integral.
% In the sequel, we continue with the general procedure and adapt it  to the various build CSs respectively to the corresponding Hilbert spaces.
%  %\section{Quantification dans le plan complexe}
%  \subsection{Quantization of elementary classical observables}
In the complex plane, the Berezin-Klauder-Toeplitz quantization of elementary classical variables $z$ and	$\bar z$ is realized via the maps $z \mapsto A_z$ and $\bar z \mapsto A_{\bar z}$ defined on the corresponding Hilbert spaces.
\subsection{Standard quantization}
The corresponding quantized expressions of the classical variables on the Hilbert space $\mathfrak{H}_e$ are obtained as follows:
%  \begin{enumerate}
%  	\item [(i)] {$z\longrightarrow A^e_z$}, we have
\begin{eqnarray}
A_z^e %&=&%%\int_{\mathbb{C}} \frac{d^2z}{\pi} |z, \gamma\rangle_e \, {}_e\langle z, \gamma| \, W_e(|z|^2, \gamma) \, z^2 \cr
%%&=&\sum_{n,k=0}^{\infty} \int_{\mathbb{C}} \frac{d^2z}{\pi} \frac{z^{2n} \overline{z}^{2k} z^2}{\sqrt{4^{2n+2k} \left( \frac{\gamma}{2} + 1 \right)_{2n} \left( \frac{\gamma}{2} + 1 \right)_{2k}}} \cdot \frac{W_e(|z|^2, \gamma)}{N_e(|z|^2, \gamma)} |2n, \gamma\rangle \langle 2k, \gamma|\cr
&=& 4 \sum_{n=0}^\infty \sqrt{\left( \frac{\gamma}{2} + 2n \right) \left( \frac{\gamma}{2} + 2n - 1 \right)} \, |2n - 2, \gamma\rangle \langle 2n, \gamma|,\nonumber\\
A_{\overline{z}}^e&=&4 \sum_{n=0}^\infty \sqrt{\left( \frac{\gamma}{2} + 2n \right) \left( \frac{\gamma}{2} + 2n - 1 \right)} \, |2n, \gamma\rangle \langle 2n-2, \gamma|,\nonumber\\
A_{|z|^2}^e&=& 4 \sum_{n=0}^\infty \left( \frac{\gamma}{2} + 2n+1 \right) \, |2n, \gamma\rangle \langle 2n, \gamma|;\nonumber
\end{eqnarray}

and on the Hilbert space $\mathfrak{H}_e$ are obtained as follows:
\begin{eqnarray}
A_z^o&=&4 \sum_{n=0}^\infty \sqrt{\left( \frac{\gamma}{2} + 2n \right) \left( \frac{\gamma}{2} + 2n + 1 \right)} \, |2n - 1, \gamma\rangle \langle 2n+1, \gamma|,\nonumber\\
A_{\overline{z}}^o						&=& 4 \sum_{n=0}^\infty \sqrt{\left( \frac{\gamma}{2} + 2n \right) \left( \frac{\gamma}{2} + 2n + 1 \right)} \, |2n + 1, \gamma\rangle \langle 2n-1, \gamma|,\nonumber\\
A_{|z|^2}^o&=& 4 \sum_{n=0}^\infty \left( \frac{\gamma}{2} + 2n+2 \right) \, |2n+1, \gamma\rangle \langle 2n+1, \gamma|.\nonumber
\end{eqnarray}
%
%  Using the polar coordinates of the complex nuober $z = |z| e^{i\theta}$,and setting $k = n$, we obtain :
%
%  \begin{eqnarray}
%  A_{|z|^2}^o &=& \sum_{n=0}^{\infty} \left( \int_0^\infty |z|^{4n+4} \widetilde{W}_o(|z|^2, \gamma) \, d(|z|^2) \right) \cdot \int_0^{2\pi} \frac{d\theta}{\pi} \cr
%  &&\times \frac{1}{\sqrt{4^{4n+4} \left( \frac{\gamma}{2} + 1 \right)_{2n+1} \left( \frac{\gamma}{2} + 1 \right)_{2n+1}}} \, |2n+1, \gamma\rangle \langle 2n+1, \gamma|\cr
%  &=& \sum_{n=0}^{\infty} \left( \int_0^\infty |z|^{4n+4} \MeijerG{2,0}{1,2}{-1}{-1;\frac{\gamma}{2}}{\frac{|z|^2}{4}} d(|z|^2) \right) \cr
%  &&\times \frac{1}{\Gamma\left(\frac{\gamma}{2}+1\right)}\int_0^{2\pi} \frac{d\theta}{\pi} \frac{1}{\sqrt{4^{4n+4} \left( \frac{\gamma}{2} + 1 \right)_{2n+1} \left( \frac{\gamma}{2} + 1 \right)_{2n+1}}} \, |2n+1, \gamma\rangle \langle 2n+1, \gamma|\cr
%  &=& \frac{4^{2n+3}\left( \frac{\gamma}{2} + 2n+2 \right)\Gamma\left( \frac{\gamma}{2} + 2n + 2 \right)}{\sqrt{4^{4n+4}\Gamma\left( \frac{\gamma}{2} + 2n + 2 \right)^2}} \, |2n+1, \gamma\rangle \langle 2n+1, \gamma|\cr
%  &=& 4 \sum_{n=0}^\infty \left( \frac{\gamma}{2} + 2n+2 \right) \, |2n+1, \gamma\rangle \langle 2n+1, \gamma|
%  \end{eqnarray}
% \end{enumerate}

Taking the  generators of the $\mathfrak{su}(1,1)$ Lie algebra and performing their expectation value given by $\langle K_{\pm} \rangle=\langle z,\gamma |K_{\pm}|z,\gamma\rangle$ or $\langle K_{3} \rangle=\langle z,\gamma |K_{3}|z,\gamma\rangle$, we get respectively to the Hilbert spaces $\mathfrak{H}_e$ and $\mathfrak{H}_o$, the following results:
%  Perform the expectation values of the corresponding Lie algebra.
%  \begin{equation}
%  \langle K_-\rangle=\langle z,\gamma |K_-|z,\gamma\rangle
%  \end{equation}
%  \begin{eqnarray}
%  && K_-|z,\gamma\rangle_e=z|z,\gamma\rangle_e\qquad K_-|z,\gamma\rangle_o=z|z,\gamma\rangle_o\\
%  &&_e\langle z,\gamma |K_+=_e\langle z,\gamma |\overline{z}\qquad_o\langle z,\gamma |K_+=_o\langle z,\gamma |\overline{z}\\
%  && K_3|z,\gamma\rangle_e=(4n+\gamma)|z,\gamma\rangle_e\qquad K_3 | z,\gamma \rangle_o=(4n+\gamma+2)| z,\gamma \rangle_o
%&& K_|z,\gamma\rangle_o=z|z,\gamma\rangle_o\qquad_o\langle z,\gamma |K_+=_o\langle z,\gamma |\overline{z}
%  \end{eqnarray}
\begin{equation*}
_e\langle K_-\rangle_e =_e\langle z,\gamma |z|z,\gamma\rangle_e=z, \quad
_e\langle K_+\rangle_e =_e\langle z,\gamma |\overline{z}|z,\gamma\rangle_e=\overline{z},
\end{equation*}
\begin{equation*}
_o\langle K_-\rangle_o=_o\langle z,\gamma |z|z,\gamma\rangle_o=z, \quad
_o\langle K_+\rangle_o=_o\langle z,\gamma |\overline{z}|z,\gamma\rangle_o=\overline{z},
\end{equation*}
\begin{equation*}
_e\langle K_3\rangle_e=_e\langle z,\gamma |4n+\gamma|z,\gamma\rangle_e=4n+\gamma, \quad
_o\langle K_3\rangle_o=_o\langle z,\gamma |4n+\gamma+2|z,\gamma\rangle_o=4n+\gamma+2.
\end{equation*}
Thereby, the relevant expectation values are provided on the Hilbert space $\mathfrak{H}_e$ as
\begin{eqnarray}
_e\langle z,\gamma|A_{\overline{z}}^e|z,\gamma\rangle_e=_e\langle z,\gamma|\overline{z}^2|z,\gamma\rangle_e=\overline{z}^2,\qquad
%\end{eqnarray}
%ù\begin{eqnarray}
_e\langle z,\gamma|A_z^e|z,\gamma\rangle_e=_e\langle z,\gamma|z^2|z,\gamma\rangle_e=z^2,\nonumber
\end{eqnarray}
\begin{eqnarray}
_e\langle z,\gamma|A_{|z|^2}\^e|z,\gamma\rangle_e=_e\langle z,\gamma|2(\gamma+4n+2)|z,\gamma\rangle_e=2(\gamma+4n+2),\nonumber
\end{eqnarray}
and on the Hilbert space $\mathfrak{H}_o$, they are given by
\begin{eqnarray}
_o\langle z,\gamma|A_{\overline{z}}^o|z,\gamma\rangle_o=_o\langle z,\gamma|\overline{z}^2|z,\gamma\rangle_o=\overline{z}^2,\qquad
%%%	\end{eqnarray}
%	\begin{eqnarray}
_o\langle z,\gamma|A_z^o|z,\gamma\rangle_o=_o\langle z,\gamma|z^2|z,\gamma\rangle_o=z^2,\nonumber
\end{eqnarray}
%  \begin{eqnarray}
%  A_{|z|^2}^o|z,\gamma\rangle_e&=&2(\gamma+4n+4)|z,\gamma\rangle_o,
%  \end{eqnarray}
\begin{eqnarray}
_o\langle z,\gamma|A_{|z|^2}^o|z,\gamma\rangle_o=_o\langle z,\gamma|2(\gamma+4n+4)|z,\gamma\rangle_o=2(\gamma+4n+2).\nonumber
\end{eqnarray}
{
	%  \color{red}
	A summary of the standard expectation values of the quantum observables of the system is given as follows:
	\begin{center}
		{\small
			\begin{tabular}{|c|l|l|l|}
				\hline
				{\bf Coherent states}&{\bf SU(1,1) generators mean values}&{\bf Quantized  observables mean values}\\
				\hline
				\multirow{2}{2.5cm}{\bf Even GBCSs }
				&$\tiny {}_e\langle K_-\rangle_e=z$&$\tiny {}_e\langle A_z^e\rangle_e=z^2$\\ \cline{2-3}
				&${}_e\langle K_+\rangle_e=\overline{z}$ &${}_e\langle A^e_{\overline{z}}\rangle_e=\overline{z}^2$ \\ \cline{2-3}
				%\cline{3-4}
				&${}_e\langle K_3\rangle_e=4n+\gamma$ &$ {}_e\langle A_{|z|^2}^e\rangle_e=2(4n+\gamma+2)$ \\ \cline{2-3}
				\hline
				\multirow{2}{2.5cm}{\bf Odd GBCSs }
				&$\tiny {}_o\langle K_-\rangle_o=z$&$\tiny {}_o\langle A_z^o\rangle_o=z^2$\\ \cline{2-3}
				&${}_o\langle K_+\rangle_o=\overline{z}$ &${}_o\langle A^o_{\overline{z}}\rangle_o=\overline{z}^2$ \\ \cline{2-3}
				&${}_o\langle K_3\rangle_o=4n+\gamma+2$ &$ {}_o\langle A_{|z|^2}^o\rangle_o=2(4n+\gamma+4)$ \\ \cline{2-3}
				\hline
		\end{tabular}}
\end{center}}

\subsection{Quantization from the DOOT technique}

Here we evaluate the quantized values of the corresponding classical observables through the DOOT technique. For each set of CSs, we obtain the following results.
%  {\color{red}
%  Then, we deduce the  expectation values in the DOOT framework respectively and compare them  to the standard ones obtained in the previous paragraph.}

%\subsubsection{Case of the even BGCSs} 

The correpondance from the classical observables to the operators leads to:
\begin{enumerate}
	\item [(i)]Case of the even BGCSs
	\begin{eqnarray}
	A_z^e
	%&=&\sum_{n=0}^{\infty}\sqrt{4(\gamma+4n)(\gamma+4n-2)}|2n-2,\gamma\rangle\langle 2n,\gamma|\cr
	%					&=&\sum_{n=0}^{\infty}\sqrt{4(\gamma+4n)(\gamma+4n-2)}\#\frac{(K_+)^{2n-2}}{\sqrt{4^{2n-2}\left(\frac{\gamma}{2}+1\right)_{2n-2}}}|0,\gamma\rangle\langle 0,\gamma|\frac{(K_-)^{2n}}{\sqrt{4^{2n}\left(\frac{\gamma}{2}+1\right)_{2n}}}\#\cr
	%					&=&\sum_{n=0}^{\infty}\sqrt{4(\gamma+4n)(\gamma+4n-2)}\#\frac{(K_+K_-)^{2n-2}K_-^2}{\sqrt{4^{4n-4}\cdot 4(\gamma+4n)(\gamma+4n-2)\left(\frac{\gamma}{2}+1\right)_{2n-2}^2}}|0,\gamma\rangle\langle 0,\gamma|\#\cr
	%					&=&K_-^2\sum_{n=0}^{\infty}\#{}_1F_1\left(1; \frac{\gamma}{2}+1; \frac{K_+K_-}{4} \right)\frac{1}{{}_1F_1\left(1; \frac{\gamma}{2}+1; \frac{K_+K_-}{4} \right)}\#\cr
	=K_-^2,\qquad
	%\end{eqnarray}
	%	\begin{eqnarray}
	A_{\overline{z}}^e%&=&\sum_{n=0}^{\infty}\sqrt{4(\gamma+4n)(\gamma+4n-2)}|2n,\gamma\rangle\langle 2n-2,\gamma|\cr
	%					&=&\sum_{n=0}^{\infty}\sqrt{4(\gamma+4n)(\gamma+4n-2)}\#\frac{(K_-)^{2n-2}}{\sqrt{4^{2n-2}\left(\frac{\gamma}{2}+1\right)_{2n-2}}}|0,\gamma\rangle\langle 0,\gamma|\frac{(K_+)^{2n}}{\sqrt{4^{2n}\left(\frac{\gamma}{2}+1\right)_{2n}}}\#\cr
	%					&=&\sum_{n=0}^{\infty}\sqrt{4(\gamma+4n)(\gamma+4n-2)}\#\frac{(K_+K_-)^{2n-2}K_+^2}{\sqrt{4^{4n-4}\cdot 4(\gamma+4n)(\gamma+4n-2)\left(\frac{\gamma}{2}+1\right)_{2n-2}^2}}|0,\gamma\rangle\langle 0,\gamma|\#\cr
	%					&=&K_+^2\sum_{n=0}^{\infty}\#{}_1F_1\left(1; \frac{\gamma}{2}+1; \frac{K_+K_-}{4} \right)\frac{1}{{}_1F_1\left(1; \frac{\gamma}{2}+1; \frac{K_+K_-}{4} \right)}\#\cr
	=K_+^2,\qquad
	%	\end{eqnarray}
	%\begin{eqnarray}
	A_{|z|^2}^e%&=&2\sum_{n=0}^{\infty}(\gamma+4n+2)|2n,\gamma\rangle\langle 2n,\gamma|\cr
	%		 			&=&2\sum_{n=0}^{\infty}(K_0+2)\#\frac{(K_-)^{2n}}{\sqrt{4^{2n}\left(\frac{\gamma}{2}+1\right)_{2n}}}|0,\gamma\rangle\langle 0,\gamma|\frac{(K_+)^{2n}}{\sqrt{4^{2n}\left(\frac{\gamma}{2}+1\right)_{2n}}}\#\cr
	%					&=&2\sum_{n=0}^{\infty}(K_0+2)\#\frac{(K_+K_-)^{2n}}{\sqrt{4^{4n}\left(\frac{\gamma}{2}+1\right)_{2n}^2}}|0,\gamma\rangle\langle 0,\gamma|\#\cr
	%					&=&2\left[2\sum_{n=0}^{\infty}\#_1F_1\left(1; \frac{\gamma}{2}+1; \frac{K_+K_-}{4} \right)+K_0\#{}_1F_1\left(1; \frac{\gamma}{2}+1; \frac{K_+K_-}{4} \right)\right]\cr
	%					&&\times\frac{1}{{}_1F_1\left(1; \frac{\gamma}{2}+1; \frac{K_+K_-}{4} \right)}\#\cr
	=2K_0+4\mathbb{I_{\mathfrak H}}.
	\end{eqnarray}
	\item [(ii)]Case of the odd BGCSs
	%% The quantization of the classical observables in the complex plane supplies:
	\begin{eqnarray}
	A_z^o%&=&\sum_{n=0}^{\infty}\sqrt{4(\gamma+4n)(\gamma+4n+2)}|2n-1,\gamma\rangle\langle 2n+1,\gamma|\cr
	%					&=&\sum_{n=0}^{\infty}\sqrt{4(\gamma+4n)(\gamma+4n+2)}\#\frac{(K_+)^{2n-1}}{\sqrt{4^{2n-1}\left(\frac{\gamma}{2}+1\right)_{2n-1}}}|0,\gamma\rangle\langle 0,\gamma|\frac{(K_-)^{2n+1}}{\sqrt{4^{2n+1}\left(\frac{\gamma}{2}+1\right)_{2n+1}}}\#\cr
	%					&=&\sum_{n=0}^{\infty}\sqrt{4(\gamma+4n)(\gamma+4n+2)}\#\frac{(K_+K_-)^{2n-1}K_-^2}{\sqrt{4^{4n-2}\cdot 4(\gamma+4n)(\gamma+4n+2)\left(\frac{\gamma}{2}+1\right)_{2n-1}^2}}|0,\gamma\rangle\langle 0,\gamma|\#\cr
	%					&=&K_-^2\sum_{n=0}^{\infty}\#{}_1F_1\left(1; \frac{\gamma}{2}+1; \frac{K_+K_-}{4} \right)\frac{1}{{}_1F_1\left(1; \frac{\gamma}{2}+1; \frac{K_+K_-}{4} \right)}\#\cr
	=K_-^2,\qquad
	%				\end{eqnarray}
	%				\begin{eqnarray}
	A_{\overline{z}}^o%%&=&\sum_{n=0}^{\infty}\sqrt{4(\gamma+4n)(\gamma+4n+2)}|2n+1,\gamma\rangle\langle 2n-1,\gamma|\cr
	%					&=&\sum_{n=0}^{\infty}\sqrt{4(\gamma+4n)(\gamma+4n+2)}\#\frac{(K_-)^{2n-1}}{\sqrt{4^{2n-1}\left(\frac{\gamma}{2}+1\right)_{2n-1}}}|0,\gamma\rangle\langle 0,\gamma|\frac{(K_+)^{2n+1}}{\sqrt{4^{2n+1}\left(\frac{\gamma}{2}+1\right)_{2n+1}}}\#\cr
	%					&=&\sum_{n=0}^{\infty}\sqrt{4(\gamma+4n)(\gamma+4n+2)}\#\frac{(K_+K_-)^{2n-1}K_+^2}{\sqrt{4^{4n-2}\cdot 4(\gamma+4n)(\gamma+4n+2)\left(\frac{\gamma}{2}+1\right)_{2n-1}^2}}|0,\gamma\rangle\langle 0,\gamma|\#\cr
	%					&=&K_+^2\sum_{n=0}^{\infty}\#{}_1F_1\left(1; \frac{\gamma}{2}+1; \frac{K_+K_-}{4} \right)\frac{1}{{}_1F_1\left(1; \frac{\gamma}{2}+1; \frac{K_+K_-}{4} \right)}\#\cr
	=K_+^2,\qquad
	%				\end{eqnarray}
	%				\begin{eqnarray}
	A_{|z|^2}^o%&=&2\sum_{n=0}^{\infty}(\gamma+4n+4)|2n+1,\gamma\rangle\langle 2n+1,\gamma|\cr
	%					&=&2\sum_{n=0}^{\infty}(K_0+4)\#\frac{(K_-)^{2n+1}}{\sqrt{4^{2n+1}\left(\frac{\gamma}{2}+1\right)_{2n+1}}}|0,\gamma\rangle\langle 0,\gamma|\frac{(K_+)^{2n+1}}{\sqrt{4^{2n+1}\left(\frac{\gamma}{2}+1\right)_{2n+1}}}\#\cr
	%					&=&2\sum_{n=0}^{\infty}(K_0+2)\#\frac{(K_+K_-)^{2n+1}}{\sqrt{4^{4n+2}\left(\frac{\gamma}{2}+1\right)_{2n+1}^2}}|0,\gamma\rangle\langle 0,\gamma|\#\cr
	%					&=&2\left[4\sum_{n=0}^{\infty}\#_1F_1\left(1; \frac{\gamma}{2}+1; \frac{K_+K_-}{4} \right)+K_0\#{}_1F_1\left(1; \frac{\gamma}{2}+1; \frac{K_+K_-}{4} \right)\right]\cr
	%					&&\times\frac{1}{{}_1F_1\left(1; \frac{\gamma}{2}+1; \frac{K_+K_-}{4} \right)}\#\cr
	=2K_0+8\mathbb{I_{\mathfrak H}}
	\end{eqnarray}
\end{enumerate}

\section{Density operator expansion in the constructed CSs basis
	% Thermal  properties of the different classes of CSs
}
% 
% If we consider that the whole quantum system (a + b) obeys
% the canonical distribution, then a mixed state, in which
% both individual parts have an equal probability distribution
% function, is characterized by the following density operator:
% 
% \\

\subsection{Density operator construction using the DOOT technique}
In quantum mechanics, the probability distribution over the states of a physical system can be characterized by a statistical operator, called the density operator, when the entire quantum system is assumed to follow a canonical distribution. When a system is in thermal contact with a reservoir (or \textquotedblleft thermal bath\textquotedblright), its state becomes mixed, described by the canonical equilibrium normalized density operator, usually denoted  $\rho$, and given by
\begin{equation}
\rho=\frac{1}{Z}\sum_{n=0}^{\infty}e^{-\beta e_n}|n\rangle\langle n|
\end{equation}
where the partition function $Z$  is taken as the normalization constant, with $\beta = \frac{1}{k_B T}$, $k_B$ being the Boltzmann constant and $T$ the system  environment temperature. In the following, we will apply the DOOT (Dynamic Operator Ordering Transformation) operation to the canonical density operator. The corresponding density operators and related partition functions are obtained in the Fock Hilbert space, using Eqs.(\ref{Over1}) and (\ref{over003}) together as key ingredients. We then successively obtain, for each set of CSs:
\begin{enumerate}
	\item For the even  Barut-Girardello CSs, 
	\begin{equation*}
	\rho_e^{\gamma}
	= \frac{1}{Z_e}\sum_{n=0}^{\infty}e^{-8n\beta}e^{-2\beta\gamma}|2n,\gamma\rangle\langle 2n,\gamma|,\qquad Z_e  = \frac{e^{-2\beta\gamma}}{1-e^{-8\beta}}, 
	\end{equation*} 
	with the DOOT transformed density operator given by
	\begin{equation}\label{dootdens001}
	\rho_e^{\gamma}
	= (1-e^{-8\beta})\#\frac{_1F_1\left(1;\frac{\gamma}{2}+1;\frac{e^{-4\beta}K_+K_-}{4}\right)}{_1F_1\left(1;\frac{\gamma}{2}+1;\frac{K_+K_-}{4}\right)}\#.
	\end{equation}
	\item For the odd Barut-Girardello CSs, 
	\begin{equation*}
	\rho_o^{\gamma}
	= \frac{1}{Z_o}\sum_{n=0}^{\infty}e^{-(8n+4)\beta}e^{-2\beta\gamma}|2n+1,\gamma\rangle\langle 2n+1,\gamma|, \qquad Z_o = \frac{e^{-2\beta(\gamma+1)}}{1-e^{-8\beta}}, 
	\end{equation*}
	with the DOOT counterpart given by
	\begin{equation}\label{dootdens003}
	\rho_o^{\gamma}
	=(1-e^{-8\beta})\#\frac{_1F_1\left(1;\frac{\gamma}{2}+1;\frac{e^{-4\beta}K_+K_-}{4}\right)-1}{_1F_1\left(1;\frac{\gamma}{2}+1;\frac{K_+K_-}{4}\right)-1}\#.
	\end{equation}
	\item For the Gazeau-Klauder CSs,
	\begin{equation*}
	\rho_{GKCS}^{\gamma}
	=
	\frac{1}{Z_{GKCS}}\sum_{n=0}^{\infty}e^{-4n\beta}e^{-2\beta\gamma}|\Phi_n^{\gamma}\rangle\langle \Phi_n^{\gamma}|, \qquad Z_{GKCS} 
	= 
	\frac{e^{-2\beta\gamma}}{1-e^{-4\beta}}, 
	\end{equation*}
	with the DOOT correspondent given by  
	\begin{equation}\label{dootdens005}
	\rho_{GKCS}^{\gamma}=
	(1-e^{-4\beta})\#\frac{_1F_1\left(1;\frac{\gamma}{2}+1;\frac{e^{-4\beta}K_+K_-}{4}\right)}{_1F_1\left(1;\frac{\gamma}{2}+1;\frac{K_+K_-}{4}\right)}
	\#.
	\end{equation}
\end{enumerate}
\subsection{Thermal expectations from the DOOT scheme}
This paragraph applies the above results of the density operator calculation to derive the correspondent thermal mean values of $\mathfrak{su}(1,1)$ Lie algebra generators,   and of the quantized classical observable products,  respectively, in the DOOT framework. The quantum expectation value of the $s$-th power of the product of the operators $K_+$ and $K_-,$ (or $A_z$ and $A_{\overline{z}},$) is calculated as presented below :
\begin{enumerate}
	\item [(a)] even BGCSs basis : we obtain
	{\small\begin{eqnarray}
		_e\langle\# (K_-K_+)^s\#\rangle_e&=&Tr(\#\rho_e(K_-K_+)^s\#)\cr
		&=&2(4e^{4\beta})^{s}\sinh(e^{4\beta})\frac{\Gamma\left(\frac{\gamma}{2}+s+1\right)}{\Gamma\left(\frac{\gamma}{2}+1\right)}\,_2F_1\left(1,\frac{\gamma}{2}+s+1; \frac{\gamma}{2}+1;e^{4\beta} \right).
		\end{eqnarray}}
	For the quantized classical observables, we get
	{\small\begin{eqnarray}
		_e\langle\# (A_z^e A_{\overline{z}}^e)^s\#\rangle_e&=&Tr(\#\rho_e(A_z^eA_{\overline{z}}^e)^s\#)\cr
		&=&2(4e^{4\beta})^{2s}\sinh(e^{4\beta})\frac{\Gamma\left(\frac{\gamma}{2}+2s+1\right)}{\Gamma\left(\frac{\gamma}{2}+1\right)}\,_2F_1\left(1,\frac{\gamma}{2}+2s+1; \frac{\gamma}{2}+1;e^{4\beta} \right);
		\end{eqnarray}}
	\item [(b)]odd BGCSs basis: we obtain the following results
	{\small\begin{eqnarray}
		_o\langle\# (K_-K_+)^s\#\rangle_o&=&Tr(\#\rho_oK_-K_+)^s\#)\cr
		&=&(4e^{4\beta})^{s+1}\sinh(e^{4\beta})\frac{\Gamma\left(\frac{\gamma}{2}+s+2\right)}{2\Gamma\left(\frac{\gamma}{2}+2\right)}\,_2F_1\left(1,\frac{\gamma}{2}+s+2; \frac{\gamma}{2}+2;e^{4\beta} \right),
		\end{eqnarray}}
	{\small	\begin{eqnarray}
		_o\langle\# (A_z^o A_{\overline{z}}^o)^s\#\rangle_o&=&Tr(\#\rho_o(A_z^oA_{\overline{z}}^o)^s\#)\cr
		&=&(4e^{4\beta})^{2s+1}\sinh(e^{4\beta})\frac{\Gamma\left(\frac{\gamma}{2}+2s+2\right)}{2\Gamma\left(\frac{\gamma}{2}+2\right)}\,_2F_1\left(1,\frac{\gamma}{2}+2s+2; \frac{\gamma}{2}+2;e^{4\beta} \right).
		\end{eqnarray}}
\end{enumerate}

{
	% 		\color{red}
	\subsection{Thermal mean values from the standard density operator expansion}
	From the usual density operator expansion, the  thermal expectations of the quantum system  Lie algebra generators, and  the classical observables related operators are provided as follows.
	%
	% 	\begin{enumerate}
	%
	% 		\item Investigate the calculation of the standard thermal mean values by quantization
	%
	% 		We investigate the standard thermal expectation values of the operators $K_-$ and $K_+$ of the $su(1,1)$ group, as well as of the quantization observables $A_z$ and $A_{\overline{z}}$.
	\subsubsection{Case of the even BGCSs}
	We have
	\begin{eqnarray}
	_e\langle K_-\rangle_e&=&Tr(\rho_e K_-)\cr
	%							&=&\int_{\mathbb{C}} \frac{d^2z}{\pi} \, {}_e\langle z, \gamma| K_-|z, \gamma\rangle_e\, P_e(|z|^2, \gamma)  \, W_e(|z|^2, \gamma)\cr
	%							&=&\int_{\mathbb{C}} \frac{d^2z}{\pi} z P_e(|z|^2, \gamma)  \, W_e(|z|^2, \gamma)\cr
	&=&\int_{0}^{\infty} |z|(1-e^{-8\beta})\frac{\MeijerG{2,0}{1,2}{-1}{-1;\frac{\gamma}{2}}{\frac{|z|^2}{4}}\MeijerG{2,0}{1,2}{-1}{-1;\frac{\gamma}{2}}{\frac{e^{-4\beta}|z|^2}{4}}}{4\Gamma\left(\frac{\gamma}{2}+1\right)\MeijerG{2,0}{1,2}{-1}{-1;\frac{\gamma}{2}}{\frac{|z|^2}{4}}}\cr
	&&\times\,_1F_1\left(1; \frac{\gamma}{2}+1; \frac{|z|^2}{4}\right) d(|z|^2)\int_{0}^{2\pi}e^{-i\theta}\frac{d\theta}{\pi}.\nonumber
	\end{eqnarray}
	The angular integration leads to
	\begin{equation*}
	\int_{0}^{2\pi}e^{-i\theta}\frac{d\theta}{\pi}=0,
	\end{equation*}
	and we obtain
	\begin{equation*}
	_e\langle K_-\rangle_e=0.
	\end{equation*}
	Likewise, for the operator $K_+$, we obtain the corresponding result by analogy:
	%then, we obtain
	\begin{equation*}
	_e\langle K_+\rangle_e=0.
	\end{equation*}
	For the quantized classical observables,  we get
	\begin{eqnarray}
	_e\langle A_z\rangle_e&=&Tr(\rho_e A_z)\cr
	%							&=&\int_{\mathbb{C}} \frac{d^2z}{\pi} \, {}_e\langle z, \gamma| A_z|z, \gamma\rangle_e\, P_e(|z|^2, \gamma)  \, W_e(|z|^2, \gamma)\cr
	%							&=&\int_{\mathbb{C}} \frac{d^2z}{\pi} z^2 P_e(|z|^2, \gamma)  \, W_e(|z|^2, \gamma)\cr
	&=&\int_{0}^{\infty} |z|^2(1-e^{-8\beta})\frac{\MeijerG{2,0}{1,2}{-1}{-1;\frac{\gamma}{2}}{\frac{|z|^2}{4}}\MeijerG{2,0}{1,2}{-1}{-1;\frac{\gamma}{2}}{\frac{e^{-4\beta}|z|^2}{4}}}{4\Gamma\left(\frac{\gamma}{2}+1\right)\MeijerG{2,0}{1,2}{-1}{-1;\frac{\gamma}{2}}{\frac{|z|^2}{4}}}\cr
	&&\times\,_1F_1\left(1; \frac{\gamma}{2}+1; \frac{|z|^2}{4}\right) d(|z|^2)\int_{0}^{2\pi}e^{-2i\theta}\frac{d\theta}{\pi}.\nonumber
	\end{eqnarray}
	The angular integration leads to
	\begin{equation*}
	\int_{0}^{2\pi}e^{-2i\theta}\frac{d\theta}{\pi}=0,
	\end{equation*}
	and we obtain
	\begin{equation*}
	_e\langle A_z\rangle_e=0.
	\end{equation*}
	Likewise, for the operator $A_{\overline{z}}$, we obtain the corresponding result by analogy:
	%	then, we obtain
	\begin{equation*}
	_e\langle A_{\overline{z}}\rangle_e=0.
	\end{equation*}
	\subsubsection{Case of the odd BGCSs} We obtain
	\begin{eqnarray}
	_o\langle K_-\rangle_o&=&Tr(\rho_o K_-)\cr
	%							&=&\int_{\mathbb{C}} \frac{d^2z}{\pi} \, {}_o\langle z, \gamma| K_-|z, \gamma\rangle_o\, P_o(|z|^2, \gamma)  \, W_o(|z|^2, \gamma)\cr
	%							&=&\int_{\mathbb{C}} \frac{d^2z}{\pi} z P_o(|z|^2, \gamma)  \, W_o(|z|^2, \gamma)\cr
	&=&\int_{0}^{\infty} |z|(1-e^{-8\beta})\frac{\MeijerG{2,0}{1,2}{-1}{-1;\frac{\gamma}{2}}{\frac{|z|^2}{4}}\MeijerG{2,0}{1,2}{-1}{-1;\frac{\gamma}{2}}{\frac{e^{-4\beta}|z|^2}{4}}}{4\Gamma\left(\frac{\gamma}{2}+1\right)\MeijerG{2,0}{1,2}{-1}{-1;\frac{\gamma}{2}}{\frac{|z|^2}{4}}}\cr
	&&\times\,_1F_1\left(1; \frac{\gamma}{2}+1; \frac{|z|^2}{4}\right) d(|z|^2)\int_{0}^{2\pi}e^{-i\theta}\frac{d\theta}{\pi}\nonumber
	\end{eqnarray}
	The angular integration leads to
	\begin{equation*}
	\int_{0}^{2\pi}e^{-i\theta}\frac{d\theta}{\pi}=0,
	\end{equation*}
	and we get
	\begin{equation*}
	_o\langle K_-\rangle_o=0.
	\end{equation*}
	Likewise, for the operator $K_+$, we obtain the corresponding result by analogy:
	%then, we obtain
	\begin{equation*}
	_o\langle K_+\rangle_o=0.
	\end{equation*}
	Besides, we also have
	\begin{eqnarray}
	_o\langle A_z^o\rangle_o&=&Tr(\rho_o A_z^o)\cr
	%							&=&\int_{\mathbb{C}} \frac{d^2z}{\pi} \, {}_o\langle z, \gamma| A_z^o|z, \gamma\rangle_o\, P_o(|z|^2, \gamma)  \, W_o(|z|^2, \gamma)\cr
	%							&=&\int_{\mathbb{C}} \frac{d^2z}{\pi} z^2 P_o(|z|^2, \gamma)  \, W_o(|z|^2, \gamma)\cr
	&=&\int_{0}^{\infty} |z|^2(1-e^{-8\beta})\frac{\MeijerG{2,0}{1,2}{-1}{-1;\frac{\gamma}{2}}{\frac{|z|^2}{4}}\MeijerG{2,0}{1,2}{-1}{-1;\frac{\gamma}{2}}{\frac{e^{-4\beta}|z|^2}{4}}}{4\Gamma\left(\frac{\gamma}{2}+1\right)\MeijerG{2,0}{1,2}{-1}{-1;\frac{\gamma}{2}}{\frac{|z|^2}{4}}}\cr
	&&\times\,_1F_1\left(1; \frac{\gamma}{2}+1; \frac{|z|^2}{4}\right) d(|z|^2)\int_{0}^{2\pi}e^{-2i\theta}\frac{d\theta}{\pi}\nonumber
	\end{eqnarray}
	The angular integration leads to
	\begin{equation*}
	\int_{0}^{2\pi}e^{-2i\theta}\frac{d\theta}{\pi}=0,
	\end{equation*}
	and we obtain
	\begin{equation*}
	_o\langle A_z^o\rangle_o=0.
	\end{equation*}
	Likewise, for the operator $A_{\overline{z}}^o$, we obtain the corresponding result by analogy:
	%then, we obtain
	\begin{equation*}
	_o\langle A_{\overline{z}}\rangle_o=0.
	\end{equation*}
	%
	% Investigate the calculation of the standard thermal mean values by quantization
	%
	% 		We investigate the standard thermal expectation values of the combined operators $K_-$ and $K_+$ of the $su(1,1)$ group, as well as of the quantization observables $A_z$ and $A_{\overline{z}}$.
	\subsubsection{Thermal mean values of the operator products}				
	We also provide in this paragraph the thermal expectation values of product of operators from the standard density operator expansion for the various sets of BGCSs.  We have
	\begin{enumerate}
		\item even BGCSs
		{\small\begin{eqnarray}
			_e\langle (K_-K_+)^s\rangle_e&=&Tr(\rho_e(K_-K_+)^s)\cr
			&=&2(4e^{4\beta})^{s}\sinh(e^{4\beta})\frac{\Gamma\left(\frac{\gamma}{2}+s+1\right)}{\Gamma\left(\frac{\gamma}{2}+1\right)}\,_2F_1\left(1,\frac{\gamma}{2}+s+1; \frac{\gamma}{2}+1;e^{4\beta} \right),\\
			%	\end{eqnarray}
			%	\begin{eqnarray}
			_e\langle (A_z^e A_{\overline{z}}^e)^s\rangle_e&=&Tr(\rho_e(A_z^eA_{\overline{z}}^e)^s)\cr
			&=&2(4e^{4\beta})^{2s}\sinh(e^{4\beta})\frac{\Gamma\left(\frac{\gamma}{2}+2s+1\right)}{\Gamma\left(\frac{\gamma}{2}+1\right)}\,_2F_1\left(1,\frac{\gamma}{2}+2s+1; \frac{\gamma}{2}+1;e^{4\beta} \right)
			\end{eqnarray}}
		\item odd BGCSs
		{\small\begin{eqnarray}
			_o\langle (K_-K_+)^s\rangle_o&=&Tr(\rho_oK_-K_+)^s)\cr
			&=&(4e^{4\beta})^{s+1}\sinh(e^{4\beta})\frac{\Gamma\left(\frac{\gamma}{2}+s+2\right)}{2\Gamma\left(\frac{\gamma}{2}+2\right)}\,_2F_1\left(1,\frac{\gamma}{2}+s+2; \frac{\gamma}{2}+2;e^{4\beta} \right),\\
			%	\end{eqnarray}
			%	\begin{eqnarray}
			_o\langle (A_z^o A_{\overline{z}}^o)^s\rangle_o&=&Tr(\rho_o(A_z^oA_{\overline{z}}^o)^s)\cr
			&=&(4e^{4\beta})^{2s+1}\sinh(e^{4\beta})\frac{\Gamma\left(\frac{\gamma}{2}+2s+2\right)}{2\Gamma\left(\frac{\gamma}{2}+2\right)}\,_2F_1\left(1,\frac{\gamma}{2}+2s+2; \frac{\gamma}{2}+2;e^{4\beta} \right)
			\end{eqnarray}}
	\end{enumerate}
	
	%	\subsection{Brief summary of the main results}
	The summary of all  investigated mean values is as follows:
	%\begin{enumerate}
	\paragraph{(i)} For the even BGCSs
	
	%\begin{center}
	{\tiny
		\begin{tabular}{|c|l|l|l|}
			\hline
			Entity& $\mathfrak su(1,1)$ generators products% mean values
			&Quantized   observables products %mean values
			\\
			\hline
			{\bf DOOT thermal mean values}
			&$\tiny%%%{}_e\langle\#(K_-K_+)^S\#\rangle_e=
			\varDelta(1)\,_2F_1\left(.,\dots+s+1;.;. \right)$ &$\tiny %%{}_e\langle\#( A_z A_{\overline{z}})^S\#\rangle_e=
			\varDelta(2)\,_2F_1\left(.,\dots+2s+1;.;.  \right)$\\ \hline
			{\bf Standard thermal mean values}
			&$\tiny%{}_e\langle(K_-K_+)^S\rangle_e=
			\varDelta(1)\,_2F_1\left(.,\dots+s+1;.;. \right)$ &$\tiny %{}_e\langle( A_z A_{\overline{z}})^S\rangle_e=
			\varDelta(2)\,_2F_1\left(.,\dots+2s+1;.;.  \right)$\\ \hline
	\end{tabular}}
	%\end{center}
	
	where $\,_2F_1\left(.,\dots+l\,s+1;. ; . \right)=\,_2F_1\left(1,\frac{\gamma}{2}+l\,s+1; \frac{\gamma}{2}+1;e^{4\beta} \right)$ and
	
	$\varDelta(l)=2(4e^{4\beta})^{l\,s}\sinh(e^{4\beta})\frac{\Gamma\left(\frac{\gamma}{2}+l\,s+1\right)}{\Gamma\left(\frac{\gamma}{2}+1\right)},$ $\qquad l\in \mathbb{N}$
	\paragraph{(ii)} Odd BGCSs
	
	%	\begin{center}
	{\tiny
		\begin{tabular}{|c|l|l|l|}
			\hline
			Entity&$su(1,1)$ generators products %mean values
			&Quantized   observables products %mean values
			\\
			\hline
			{\bf DOOT thermal mean values}
			&$\tiny%{}_o\langle\#( K_-K_+)^S\#\rangle_o=
			\varDelta(1)\,_2F_1\left(.,\dots+s+2;.;. \right)$ &$\tiny %{}_o\langle\#( A_z A_{\overline{z}})^S\#\rangle_o=
			\varDelta(2)\,_2F_1\left(.,\dots+2s+2;.;.  \right)$\\ \hline
			{\bf Standard thermal mean values}
			&$\tiny%{}_o\langle( K_-K_+)^S\rangle_o=
			\varDelta(1)\,_2F_1\left(.,\dots+s+2;.;. \right)$ &$\tiny %{}_o\langle( A_z A_{\overline{z}})^S\rangle_o=
			\varDelta(2)\,_2F_1\left(.,\dots+2s+2;.;.  \right)$\\ \hline
	\end{tabular}}
	%	\end{center}
	
	where $\,_2F_1\left(.,\dots+l\,s+2;. ; . \right)=\,_2F_1\left(1,\frac{\gamma}{2}+l\,s+2; \frac{\gamma}{2}+1;e^{4\beta} \right)$ and
	
	$\varDelta(l)=2(4e^{4\beta})^{l\,s}\sinh(e^{4\beta})\frac{\Gamma\left(\frac{\gamma}{2}+l\,s+2\right)}{\Gamma\left(\frac{\gamma}{2}+1\right)},$ $\qquad l\in \mathbb{N}$
	%				\end{enumerate}
}
\subsection{Husimi distribution }
The Husimi function \cite{husimi} as a function in the phase space, in general viewed as Gaussian  smoothing of the Wigner function, 
is derived by use of the expected value of the density operator in a basis of CS \cite{Klauder-Skagerstam}.  
The $Q$-Husimi function is in fact  the diagonal element of the density operator in the CS representation. It can be connected with the 
normalized 
$P$-quasidistribution  function as follows
\begin{equation*}
Q(|\xi|^2) =  \langle \xi| \rho |\xi\rangle =   \int_{\mathcal D} d\nu  |\langle \xi|z \rangle|^{2} P(|z|^2),
\quad   {\int_{\mathcal D} d\nu \; Q(|\xi|^2) = 1}
\end{equation*}
with $|\xi\rangle  (\xi \in \mathbb C)$  as in \ref{covcs}.  The normalization of the density operator in the CSs (\ref{covcs}) $\{|z\rangle\}$ basis  leads to
\begin{equation}\label{thermal02}
{\mbox{Tr} \rho  =  \int_{\mathcal D}  d\nu  \, \langle z|  \rho |z\rangle =  1.}
\end{equation}
The $Q$-Husimi function is used in  information-theoretical entropy of 
some quantum oscillators, in Fisher's and 
Shannon information measures  in statistical mechanics \cite{curilef}. The diagonal elements of the density operators in the basis of the constructed CSs are called the Husimi distribution denoted $Q_{|z\rangle}$ and defined as
\begin{equation*}
Q_{|z\rangle}=\langle z|\rho |z\rangle.
\end{equation*}
For the constructed states, we will determine in the sequel the Husimi distribution in the DOOT framework.
\begin{itemize}
	\item [(a)] Even BGCSs
	%\begin{enumerate}
	%%	\item Even
	\begin{eqnarray}
	Q_{|z,\gamma\rangle_e}%&=&_e\langle z,\gamma|\rho_e |z,\gamma\rangle_e\cr
	%	&=&_e\langle z,\gamma|(1-e^{-8\beta})\#\frac{_1F_1\left(1;\frac{\gamma}{2}+1;\frac{e^{-4\beta}K_+K_-}{4}\right)}{_1F_1\left(1;\frac{\gamma}{2}+1;\frac{K_+K_-}{4}\right)}\#|z,\gamma\rangle_e\cr
	&=&(1-e^{-8\beta})\#\frac{_1F_1\left(1;\frac{\gamma}{2}+1;\frac{e^{-4\beta}|z|^2}{4}\right)}{_1F_1\left(1;\frac{\gamma}{2}+1;\frac{|z|^2}{4}\right)}\#.
	\end{eqnarray}
	\item [(b)] Odd BGCSs
	\begin{eqnarray}
	Q_{|z,\gamma\rangle_o}%%&=&_o\langle z,\gamma|\rho_o |z,\gamma\rangle_o\cr
	%&=&_o\langle z,\gamma|(1-e^{-8\beta})\#\frac{_1F_1\left(1;\frac{\gamma}{2}+1;\frac{e^{-4\beta}K_+K_-}{4}\right)-1}{_1F_1\left(1;\frac{\gamma}{2}+1;\frac{K_+K_-}{4}\right)-1}\#|z,\gamma\rangle_o\cr
	&=&(1-e^{-8\beta})\#\frac{_1F_1\left(1;\frac{\gamma}{2}+1;\frac{e^{-4\beta}|z|^2}{4}\right)-1}{_1F_1\left(1;\frac{\gamma}{2}+1;\frac{|z|^2}{4}\right)-1}\#.
	\end{eqnarray}
	%%\end{enumerate}
	\item [(c)] GKCSs
	\begin{eqnarray}
	Q_{|J,\alpha\rangle}%&=&\langle J,\alpha|\rho_{GKCS} |J,\alpha\rangle\cr
	%%	&=&\langle J,\alpha|(1-e^{-4\beta})\#\frac{_1F_1\left(1;\frac{\gamma}{2}+1;\frac{e^{-4\beta}K_+K_-}{4}\right)}{_1F_1\left(1;\frac{\gamma}{2}+1;\frac{K_+K_-}{4}\right)}\#|J,\alpha\rangle\cr
	&=&(1-e^{-4\beta})\#\frac{_1F_1\left(1;\frac{\gamma}{2}+1;\frac{e^{-4\beta}J}{4}\right)}{_1F_1\left(1;\frac{\gamma}{2}+1;\frac{J}{4}\right)}\#.
	\end{eqnarray}
\end{itemize}
\subsection{Glauber-Sudarshan $P$-diagonal representation}
This paragraph is devoted to the diagonal expansion of the density operator, also called Glauber-Sudarshan  $P$-diagonal representation,  in terms of the different sets of CSs  projectors. This expansion is useful in order to calculate the thermal
expectation values (thermal averages) of operators characterizing the quantum system
considered as a quantum gas in the thermodynamic equilibrium \cite{Popov_Pop, aremuaromp}. By definition, the diagonal $P$-representation is given by (see e.g. \cite{brif-aryeh})
\begin{equation}\label{rho000}
\rho=\int_{\mathbb{C}}w(|z|^2)P(|z|^2)|z\rangle \langle z|\frac{d^2z}{\pi}.
\end{equation}
In the basis of the constructed CSs, by vitue of Eq.(\ref{rho000}), we successively obtain the following expressions:
\begin{enumerate}
	\item Case of Barut-Girardello CSs
	\begin{itemize}
		\item [(a)] Even CSs. We have
		\begin{equation*}
		\rho_e^{\gamma} 
		=
		\int_{\mathbb{C}}W_e(|z|^2,\gamma)P_e(|z|^2,\gamma)|z,\gamma\rangle_e\,_e\langle z,\gamma|\frac{d^2z}{2\pi}
		\end{equation*}
		where $P_e(|z|^2,\gamma)$ is provided through the equality
		% 
		% 
		% 	then, we deduce
		% 	\begin{eqnarray}
		% 	\int_{0}^{\infty}\tilde{P_e}(|z|^2,\gamma)(|z|^2)^{2n}d(|z|^2)|&=&\frac{1}{Z_e}e^{-8n\beta}e^{-2\beta\gamma}\left(4^{2n}\frac{\gamma}{2}+1\right)_{2n},
		% 	\end{eqnarray}
		\begin{equation*}
		\tilde{P_e}(|z|^2,\gamma) =\frac{P_e(|z|^2,\gamma)|z|^{\gamma}e^{-\frac{|z|^2}{4}} (|z|^2)^{2n}}{2^{\gamma+2}\Gamma\left(\frac{\gamma}{2}+1\right)}\label{P121}
		\end{equation*}
		such that by use of the Mellin transform inside the normalization condition of $\tilde{P_e}(|z|^2,\gamma)$, 
		\begin{equation*}
		\int_{0}^{\infty}
		\tilde{P_e}(|z|^2,\gamma)(|z|^2)^{2n}d(|z|^2)| = \frac{e^{-2\beta\gamma}}{Z_e\Gamma\left(\frac{\gamma}{2}+1\right)}\int_{0}^{\infty}\MeijerG{2,0}{1,2}{-1}{-1;\frac{\gamma}{2}}{\frac{e^{-4\beta}|z|^2}{4}}(|z|^2)^{2n}d(|z|^2), 
		% 	\cr
		% 	\Longrightarrow
		% 	\tilde{P_e}(|z|^2,\gamma)&=&\frac{e^{-2\beta\gamma}}{Z_e\Gamma\left(\frac{\gamma}{2}+1\right)}\MeijerG{2,0}{1,2}{-1}{-1;\frac{\gamma}{2}}{\frac{e^{-4\beta}|z|^2}{4}}.
		\label{P122}
		\end{equation*}
		% 	From the equation (\ref{P121}), the equation (\ref{P122}) becomes
		we  get the appropriate $
		P$-distribution
		\begin{equation*}
		P_e(|z|^2,\gamma)=(1-e^{-8\beta})\frac{\MeijerG{2,0}{1,2}{-1}{- 1;\frac{\gamma}{2}}{\frac{e^{-4\beta}|z|^2}{4}}}{\MeijerG{2,0}{1,2}{-1}{-1;\frac{\gamma}{2}}{\frac{|z|^2}{4}}}.
		\end{equation*}
		\item [(b)] Odd CSs. We have
		\begin{eqnarray}
		\rho_o^{\gamma}  
		=
		\int_{\mathbb{C}}W_o(|z|^2,\gamma)P_o(|z|^2,\gamma)|z,\gamma\rangle_o\,_o\langle z,\gamma|\frac{d^2z}{2\pi}\nonumber
		\end{eqnarray}
		% 	then, we deduce
		% 	\begin{eqnarray}
		% 	\int_{0}^{\infty}\tilde{P_o}(|z|^2,\gamma)(|z|^2)^{2n+1}d(|z|^2)|&=&\frac{1}{Z_o}e^{-(8n+4)\beta}e^{-2\beta\gamma}\left(4^{2n+1}\frac{\gamma}{2}+1\right)_{2n+1},
		% 	\end{eqnarray}
		where $P_o(|z|^2,\gamma)$  is obtained by the following relation
		\begin{equation}
		\tilde{P_o}(|z|^2,\gamma)=\frac{P_o(|z|^2,\gamma)|z|^{\gamma}e^{-\frac{|z|^2}{4}} (|z|^2)^{2n+1}}{2^{\gamma+2}\Gamma\left(\frac{\gamma}{2}+1\right)}\label{P121o}
		\end{equation}
		with $\tilde{P_o}(|z|^2,\gamma)$ supplied by the following  integral formula 
		\begin{eqnarray}
		&&\int_{0}^{\infty}\tilde{P_o}(|z|^2,\gamma)(|z|^2)^{2n}d(|z|^2)|\cr &&=\frac{e^{-2\beta\gamma}}{Z_o\Gamma\left(\frac{\gamma}{2}+1\right)}\int_{0}^{\infty}\MeijerG{2,0}{1,2}{-1}{-1;\frac{\gamma}{2}}{\frac{e^{-4\beta}|z|^2}{4}} (|z|^2)^{2n}d(|z|^2).\nonumber
		% 	\cr	\Longrightarrow\tilde{P_o}(|z|^2,\gamma)&=&\frac{e^{-2\beta\gamma}}{Z_o\Gamma\left(\frac{\gamma}{2}+1\right)}\MeijerG{2,0}{1,2}{-1}{-1;\frac{\gamma}{2}}{\frac{e^{-4\beta}|z|^2}{4}}.
		\label{P122o}
		\end{eqnarray}
		% 	From the equation (\ref{P121o}), the equation (\ref{P122o}) becomes
		Thereby, 
		\begin{equation*}
		P_o(|z|^2,\gamma)=(1-e^{-8\beta})\frac{\MeijerG{2,0}{1,2}{-1}{-1;\frac{\gamma}{2}}{\frac{e^{-4\beta}|z|^2}{4}}}{\MeijerG{2,0}{1,2}{-1}{-1;\frac{\gamma}{2}}{\frac{|z|^2}{4}}}.
		\end{equation*}
	\end{itemize}
	\item Case of Gazeau-Klauder CSs.
	
	Following the steps as above, the definition of the diagonal  $P$-representation provides
	\begin{equation*}
	\rho_{GKCS}
	=
	\int_{\mathbb{C}}W_{GKCS}(|z|^2,\gamma)P_{GKCS}(J^2,\gamma)|J,\alpha\rangle\langle J,\alpha|dJd\alpha
	\end{equation*}
	where the use of the  Mellin transform  achieved through the equality
	\begin{equation}
	\int_{0}^{\infty}\overbrace{P_{GKCS}}(J,\gamma)J^{n}dJ=\frac{e^{-2\beta\gamma}}{Z_{GKCS}\Gamma\left(\frac{\gamma}{2}+1\right)}\int_{0}^{\infty}\MeijerG{2,0}{1,2}{-1}{-1;\frac{\gamma}{2}}{\frac{e^{-4\beta}J2}{4}}J^{n}J
	% 	\cr
	% 	\Longrightarrow\tilde{P_{GKCS}}(J,\gamma)&=&\frac{e^{-2\beta\gamma}}{Z_{GKCS}\Gamma\left(\frac{\gamma}{2}+1\right)}\MeijerG{2,0}{1,2}{-1}{-1;\frac{\gamma}{2}}{\frac{e^{-4\beta}J}{4}}.
	\label{P122g}
	\end{equation}
	provides, with the relation between both distributions $P_{GKCS}(J,\gamma)$ and $\overbrace{P_{GKCS}}(J,\gamma)$ \begin{equation}
	\overbrace{P_{GKCS}}(J,\gamma)=\frac{P_{GKCS}(J,\gamma)J|^{\gamma}e^{-\frac{J}{4}} J^{n}}{2^{\gamma+2}\Gamma\left(\frac{\gamma}{2}+1\right)}\label{P121g},
	\end{equation}
	the quasi-distribution
	\begin{equation}
	P_{GKCS}(J,\gamma)=(1-e^{-4\beta})\frac{\MeijerG{2,0}{1,2}{-1}{-1;\frac{\gamma}{2}}{\frac{e^{-4\beta}J}{4}}}{\MeijerG{2,0}{1,2}{-1}{-1;\frac{\gamma}{2}}{\frac{J}{4}}}
	\end{equation}
	as expected. 
\end{enumerate}
%
% 					\newpage

\section{Concluding remarks}
In this work, we revisited the properties of the Barut-Girardello coherent states (BGCSs) and the Gazeau-Klauder coherent states (GKCSs) using an operator ordering approach introduced in \cite{Popov-popov-miodrag}, namely the diagonal ordering operation technique (DOOT). As established in \cite{Popov-popov-miodrag, Popov-Negrea-Popov}, this approach essentially consists of applying the ordered product or operatorial functions thereof to various objects associated with the coherent states, in order to simplify the corresponding algebraic calculations. The results obtained are consistent with those derived using purely algebraic methods, thereby confirming the validity of the DOOT approach.

Throughout this paper, we applied the DOOT procedure to construct coherent states for the isotonic oscillator, starting from its discrete spectrum, and developed the associated BGCSs and GKCSs on the appropriate Hilbert spaces. We first verified that these states satisfy all of Klauder's required criteria. From the resolution of the identity property, several relevant mathematical structures were analyzed, including, in particular, the reproducing kernel properties of the constructed families of coherent states.

Once the expectation values of observables describing the quantum system were determined, several important physical properties were studied, such as the photon number distribution and the probability density. Moreover, the quantization procedure was implemented for the classical variables of the quantum system in the complex plane, both in the standard framework and within the DOOT scheme.

Subsequently, the thermal behavior of the physical system in the constructed coherent states was analyzed. The properties of mixed states described by a canonical density operator were investigated, and the corresponding density operators for each class of coherent states were computed within the DOOT framework. Relevant thermal expectation values were obtained in both the standard and DOOT treatments. Additionally, the Husimi distribution was determined, and the corresponding Glauber-Sudarshan $P$-representation was provided in the respective coherent state Hilbert spaces.

Overall, the DOOT procedure facilitates numerous calculations compared to purely algebraic methods, as evidenced by the various computed quantities, thereby confirming the observations reported in \cite{Popov-popov-miodrag, Popov-Negrea-Popov, Popov-Dong-miodrag}.

%
%
%
% \newpage
%
%
% We have defined two arbitrary partner hermitic conjugates
% densely defined and dimensionless lowering Aâ and raising
% A+ operators (which, generally, are not identical to the group
% generators of the corresponding quantum group), the only re-
% striction being that they are connected with the Hamiltonian of
% the examined system by the relation
% % H = hÌÏA+ Aâ .
% Regarding the GK-CSs introduced in Ref. [4], they can
% be obtained by applying the DOOT ordered exponential oper-
% ator
% % # exp (âiÎ³A+ Aâ ) #
% on the CSs which depend on the real
% positive variable |Ji and by considering the rules of DOOT.
% In our opinion, the main conclusions of the present paper
% are as follows.
% i) By using the new approach DOOT, we implicitly prove
% once again the validity and consistency of the elegant Gazeau
% and Klauder construction for coherent states of some quantum
% systems with a known discrete dimensionless spectrum en .
% ii) The applicability of the DOOT was extended also to
% the GK-CSs, as well as to the other (BarutâGirardello) kind
% of coherent states for many kinds of simple physical sys-
% tems (one-dimensional quantum oscillator (HO-1D), pseudo-
% harmonic oscillator (PHO), Morse oscillator, infinite quantum
% well). [18,19]
% iii) DOOT facilitates multiple calculations compared to
% purely algebraic methods.
% iv) The present results, which are identical to those ob-
% tained by using other purely algebraic methods, are demon-
% strated that the DOOT may be applied to the systems with in-
%
%

\appendix			
\section{Appendix}
{\it Proof of Proposition \ref{ident000}}. We have 
\begin{eqnarray}
\mathbb{I_{\mathfrak H}}&=&\int_{\mathbb{C}} \frac{d^2z}{\pi}W_e(|z|^2,\gamma)\ket{z,\gamma}_e\,_e\bra{z,\gamma}\cr
&=&\int_{0}^{\infty}\tilde{W_e}(|z|^2,\gamma)d(|z|^2)\int_{0}^{2\pi}\#\frac{_1F_1\left(1;\frac{\gamma}{2}+1;\frac{\overline{z}K_-}{4}\right)\,_1F_1\left(1;\frac{\gamma}{2}+1;\frac{zK_+}{4}\right)}{_1F_1\left(1;\frac{\gamma}{2}+1;\frac{K_+K_-}{4}\right)}\#\frac{d\theta}{2\pi}\label{wei12}
\end{eqnarray}
with
\begin{equation}\label{weight00}
\tilde{W_e}=\frac{W_e(|z|^2,\gamma)}{_1F_1\left(1;\frac{\gamma}{2}+1;\frac{|z|^2}{4}\right)}.
\end{equation}
Let us put the angular part as follows
\begin{equation*}
A_e(z)=\int_{0}^{2\pi}\#\frac{_1F_1\left(1;\frac{\gamma}{2}+1;\frac{\overline{z}K_-}{4}\right)\,_1F_1\left(1;\frac{\gamma}{2}+1;\frac{zK_+}{4}\right)}{_1F_1\left(1;\frac{\gamma}{2}+1;\frac{K_+K_-}{4}\right)}\#\frac{d\theta}{2\pi},
\end{equation*}
and integrate this part
\begin{equation*}
A_e(z)=\int_{0}^{2\pi}\#\frac{\sum_{n=0}^{\infty}\frac{(zK_+)^{2n}(\overline{z}K_-)^{2n}}{(4^{2n})^2\left(\frac{\gamma}{2}+1\right)_{2n}^2}}{_1F_1\left(1;\frac{\gamma}{2}+1;\frac{K_+K_-}{4}\right)}\#\frac{d\theta}{2\pi}=\sum_{n=0}^{\infty}\frac{(|z|^2)^{2n}}{4^{2n}\left(\frac{\gamma}{2}+1\right)_{2n}}|2n,\gamma\rangle\langle 2n,\gamma|.
\end{equation*}
Then, (\ref{wei12}) performs as follows
\begin{equation*}
0=\sum_{n=0}^{\infty}\left(\int_{0}^{\infty}\tilde{W_e}(|z|^2,\gamma)d(|z|^2)\frac{(|z|^2)^{2n}}{4^{2n}\left(\frac{\gamma}{2}+1\right)_{2n}}-1\right)|2n,\gamma\rangle\langle 2n,\gamma|.\label{identy1}
\end{equation*}
Applying the inverse Mellin transform to the expression in the parenthesis of (\ref{ident0001}), we  obtain the weight function as follows:
\begin{eqnarray}
\int_{0}^{\infty}\tilde{W_e}(|z|^2,\gamma)d(|z|^2)\frac{(|z|^2)^{2n}}{4^{2n}\left(\frac{\gamma}{2}+1\right)_{2n}}&=&1\cr
&=&\frac{1}{\Gamma\left(\frac{\gamma}{2}+1\right)}\int_{0}^{\infty}\MeijerG{2,0}{1,2}{-1}{-1;\frac{\gamma}{2}}{\frac{|z|^2}{4}}d(|z|^2).\label{weight2}
\end{eqnarray}
We deduce the  weight function from (\ref{weight00}) and (\ref{weight2}) as follows
\begin{eqnarray}
\tilde{W_e}(|z|^2,\gamma)&=&\frac{1}{\Gamma\left(\frac{\gamma}{2}+1\right)}\MeijerG{2,0}{1,2}{-1}{-1;\frac{\gamma}{2}}{\frac{|z|^2}{4}}=\frac{\Gamma\left(\frac{\gamma}{2}+1\right)}{\Gamma\left(\frac{\gamma}{2}+1\right)}\frac{|z|^{\gamma}e^{-\frac{|z|^2}{4}}}{2^{\gamma+2}}\cr
\Longrightarrow W_e(|z|^2,\gamma)&=&\frac{|z|^{\gamma}e^{-\frac{|z|^2}{4}}}{2^{\gamma+2}}\,_1F_1\left(1;\frac{\gamma}{2}+1;\frac{|z|^2}{4}\right).
\end{eqnarray}	\quad$\hfill{\square}$

{\it Proof of Proposition \ref{kernel000}} : Even states.

We have 
\begin{eqnarray}
K_e(z,z')&=&\int_{\mathbb{C}} K_e(z,z") K_e(z",z')\frac{W_e(|z"|^2)d^2z"}{\pi}\cr
%	&=&\int_{\mathbb{C}} \langle z",\gamma|z,\gamma\rangle\langle z',\gamma|z",\gamma\rangle\frac{W_e(|z"|^2)d^2z"}{\pi}\cr 
&=&\int_{\mathbb{C}}\langle z",\gamma|\#\frac{_1F_1\left(1;\frac{\gamma}{2}+1;\frac{\overline{z'}K_-}{4}\right)\,_1F_1\left(1;\frac{\gamma}{2}+1;\frac{zK_+}{4}\right)}{_1F_1\left(1;\frac{\gamma}{2}+1;\frac{K_+K_-}{4}\right)}|z",\gamma\rangle\#\cr
&&\frac{W_e(|z"|^2,\gamma)d(|z|^2)}{\sqrt{_1F_1\left(1;\frac{\gamma}{2}+1;\frac{|z|^2}{4}\right)\,_1F_1\left(1;\frac{\gamma}{2}+1;\frac{|z'|^2}{4}\right)}}\frac{d^2z"}{\pi}\cr
&=&\frac{\sum_{n=0}^{\infty}\frac{(z)^{2n}(\overline{z'})^{2n}}{(4^{2n})^2\left(\frac{\gamma}{2}+1\right)_{2n}^2}\frac{1}{4\Gamma\left(\frac{\gamma}{2}+1\right)}}{\sqrt{_1F_1\left(1;\frac{\gamma}{2}+1;\frac{|z|^2}{4}\right)\,_1F_1\left(1;\frac{\gamma}{2}+1;\frac{|z'|^2}{4}\right)}}\int_{0}^{\infty}|z"|^{2n}\MeijerG{2,0}{1,2}{0}{0;\frac{\gamma}{2}}{\frac{|z|^2}{4}}d(|z"|^2)\cr
%					&=&\frac{\sum_{n=0}^{\infty}\frac{(z)^{2n}(\overline{z'})^{2n}}{4^{2n}\left(\frac{\gamma}{2}+1\right)_{2n}}}{\sqrt{_1F_1\left(1;\frac{\gamma}{2}+1;\frac{|z|^2}{4}\right)\,_1F_1\left(1;\frac{\gamma}{2}+1;\frac{|z'|^2}{4}\right)}}\cr
&=&\frac{\,_1F_1\left(1;\frac{\gamma}{2}+1;\frac{z\overline{z'}}{4}\right)}{\sqrt{_1F_1\left(1;\frac{\gamma}{2}+1;\frac{|z|^2}{4}\right)\,_1F_1\left(1;\frac{\gamma}{2}+1;\frac{|z'|^2}{4}\right)}}.
\end{eqnarray} \quad$\hfill{\square}$
%%%%%%%%%%%%%%%%%%%%%%%%%%%%%%%%%%%%%%%%%%%%%%%%%%%%%%%%%%%%%%%%%%%%%%%%%%%%%%%%%%%%%%%%%%%%%%%%%%%%%%%%%%%%%%%%%%%%%%%%%%%%%%%%%%%%%%%%%%%%%%%%%%%%%%%%%%%%%%%%%%%%%%%%
%\begin{thebibliography}{10}	\addcontentsline{toc}{chapter}{References}

\end{document}